\documentclass[lettersize,journal]{IEEEtran}
\usepackage{amsmath,amsfonts,multirow}
\usepackage{algorithmic}
\usepackage{algorithm}
\usepackage{array}
\usepackage[caption=false,font=normalsize,labelfont=sf,textfont=sf]{subfig}
\usepackage{textcomp}
\usepackage{stfloats}
\usepackage{url}
\usepackage{verbatim}
\usepackage{graphicx}
\usepackage{cite}
\usepackage{subfloat}
\usepackage{bm}
\usepackage{amssymb}
\usepackage{bbding} 
\usepackage{booktabs}
\usepackage{color}
\usepackage[implicit=false]{hyperref}

\begin{document}

\title{Low-Light Hyperspectral Image Enhancement}

\author{Xuelong~Li,~\IEEEmembership{Fellow,~IEEE}, Guanlin~Li, and Bin~Zhao

\IEEEcompsocitemizethanks{
	
	\IEEEcompsocthanksitem The authors are with the School of Artificial Intelligence, OPtics and ElectroNics (iOPEN), Northwestern Polytechnical University, Xi'an 710072, P.R. China. 
	They are also with the Key Laboratory of Intelligent Interaction and Applications (Northwestern Polytechnical University), Ministry of Industry and Information Technology, Xi'an 710072, P. R. China. This work was supported in part by the National Natural Science Foundation of China under Grant 62106183. \emph{(Corresponding author: Xuelong Li)} (E-mail: li@nwpu.edu.cn; guanguanboy@gmail.com; binzhao111@gmail.com).
	
}	

}

\maketitle

\begin{abstract}
Due to inadequate energy captured by the hyperspectral camera sensor in poor illumination conditions, low-light hyperspectral images (HSIs) usually suffer from low visibility, spectral distortion, and various noises. A range of HSI restoration methods have been developed, yet their effectiveness in enhancing low-light HSIs is constrained. This work focuses on the low-light HSI enhancement task, which aims to reveal the spatial-spectral information hidden in darkened areas. To facilitate the development of low-light HSI processing, we collect a low-light HSI (LHSI) dataset of both indoor and outdoor scenes. Based on Laplacian pyramid decomposition and reconstruction, we developed an end-to-end data-driven low-light HSI enhancement (HSIE) approach trained on the LHSI dataset. With the observation that illumination is related to the low-frequency component of HSI, while textural details are closely correlated to the high-frequency component, the proposed HSIE is designed to have two branches. The illumination enhancement branch is adopted to enlighten the low-frequency component with reduced resolution. The high-frequency refinement branch is utilized for refining the high-frequency component via a predicted mask. In addition, to improve information flow and boost performance, we introduce an effective channel attention block (CAB) with residual dense connection, which served as the basic block of the illumination enhancement branch. The effectiveness and efficiency of HSIE both in quantitative assessment measures and visual effects are demonstrated by experimental results on the LHSI dataset. According to the classification performance on the remote sensing Indian Pines dataset, downstream tasks benefit from the enhanced HSI. Datasets and codes are available: \href{https://github.com/guanguanboy/HSIE}{https://github.com/guanguanboy/HSIE}.
\end{abstract}


\section{Introduction}

\noindent \IEEEPARstart{H}{yperspectral} image (HSI) is composed of substantial discrete bands for each spatial pixel. Therefore, it contains more copious information than a natural image, which benefits massive applications in HSI fusion\cite{9103204}, classification\cite{hsi_cls9091940}, \cite{8447427}, \cite{9709320}, \cite{9743446}, \cite{9565209}, remote sensing\cite{9718236}, change detection\cite{li2019unsupervised}, \cite{9444545}, visual question answering\cite{9444570}, \emph{etc.} In particular, hyperspectral imaging technology is increasingly employed for outdoor surveillance and environmental monitoring\cite{stuart2019hyperspectral}, such as real-time water quality and atmospheric pollution monitoring, which are highly significant for environmental safety. This type of full-day surveillance requires hyperspectral cameras to capture high-quality HSIs even at night. However, owing to insufficient light reaching hyperspectral camera sensors at night, the captured HSIs often affected by poor visibility, spectral distortion, and varied noises (emphasis on Gaussian, impulse, and stripe). These degradations bury the useful spatial and spectral signals of the captured HSIs, which consequently affect the performance of the aforementioned HSI applications. To alleviate such degradations, it will be beneficial to take advantage of more advanced HSI imaging hardware devices equipped with specialized photographic techniques which are not easily affordable. However, even with advanced HSI imaging devices, it is still hard to prevent the presence of noises and spectral distortion. Therefore, it is essential to design an effective algorithm to solve the low-light HSI enhancement problem. This work focuses on low-light HSI enhancement, which aims to reduce spectral distortion, suppress noises, and reveal hidden information in low-light simultaneously. 

\begin{figure}[t]
	
	\begin{minipage}[b]{0.49\textwidth}
		\centering
		\centerline{\includegraphics[width=1\textwidth, height=3.3cm]{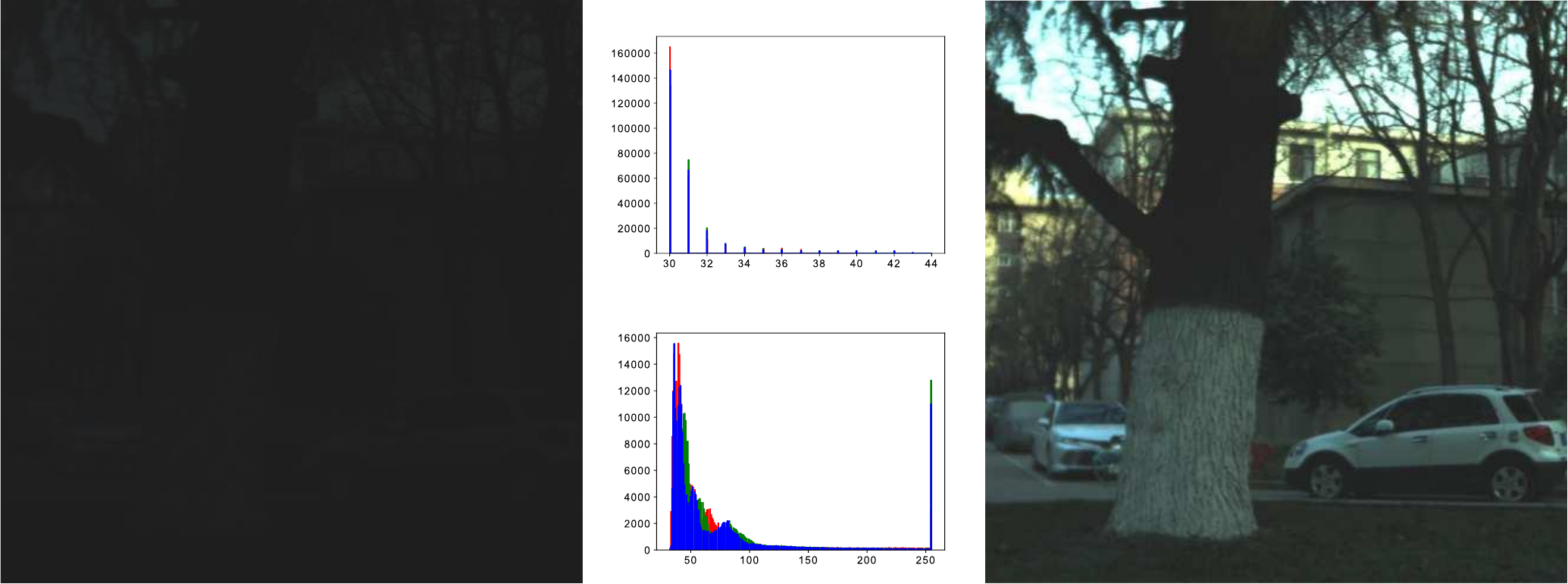}}
		\centerline{(a) Original HSIs, MSE=104.0 }\medskip
	\end{minipage}
	\hfill
	\begin{minipage}[b]{0.49\textwidth}
		\centering
		\centerline{\includegraphics[width=1\textwidth,height=3.3cm]{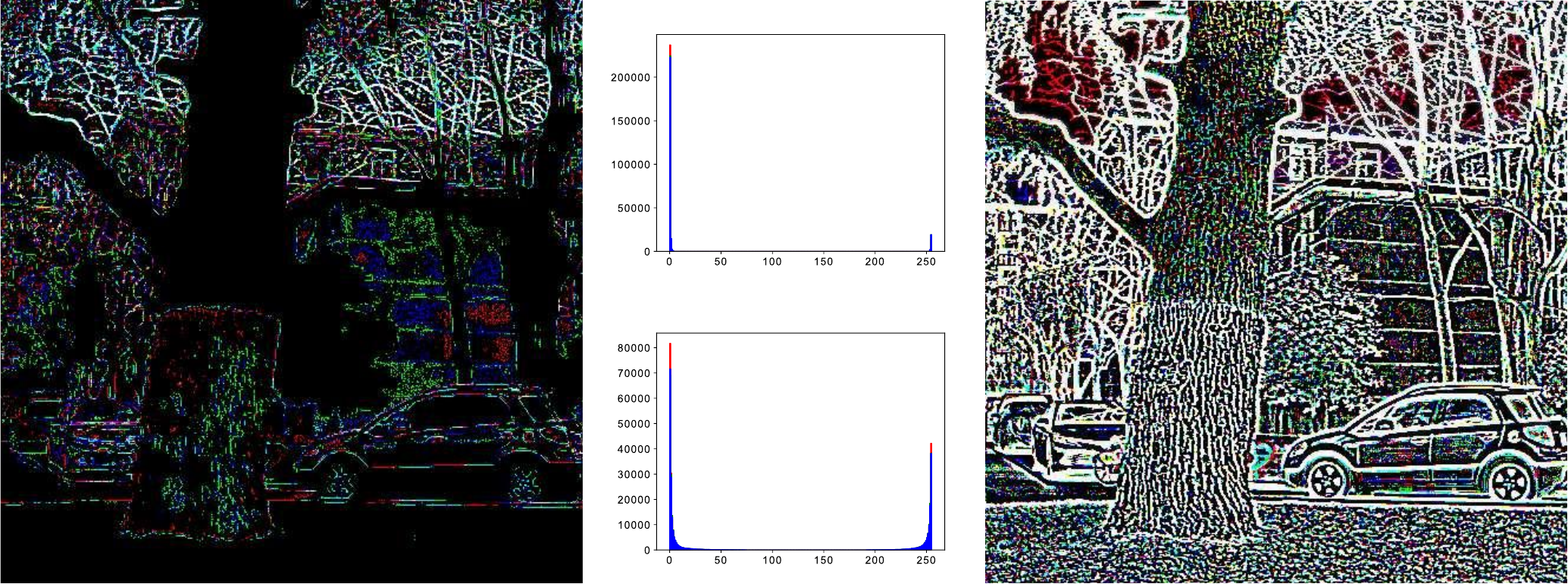}}
		\centerline{(b) High-Frequency Component, Level=1, MSE=21.6 }\medskip
	\end{minipage}
	\hfill
	\begin{minipage}[b]{0.49\textwidth}
		\centering
		\centerline{\includegraphics[width=1\textwidth,height=3.3cm]{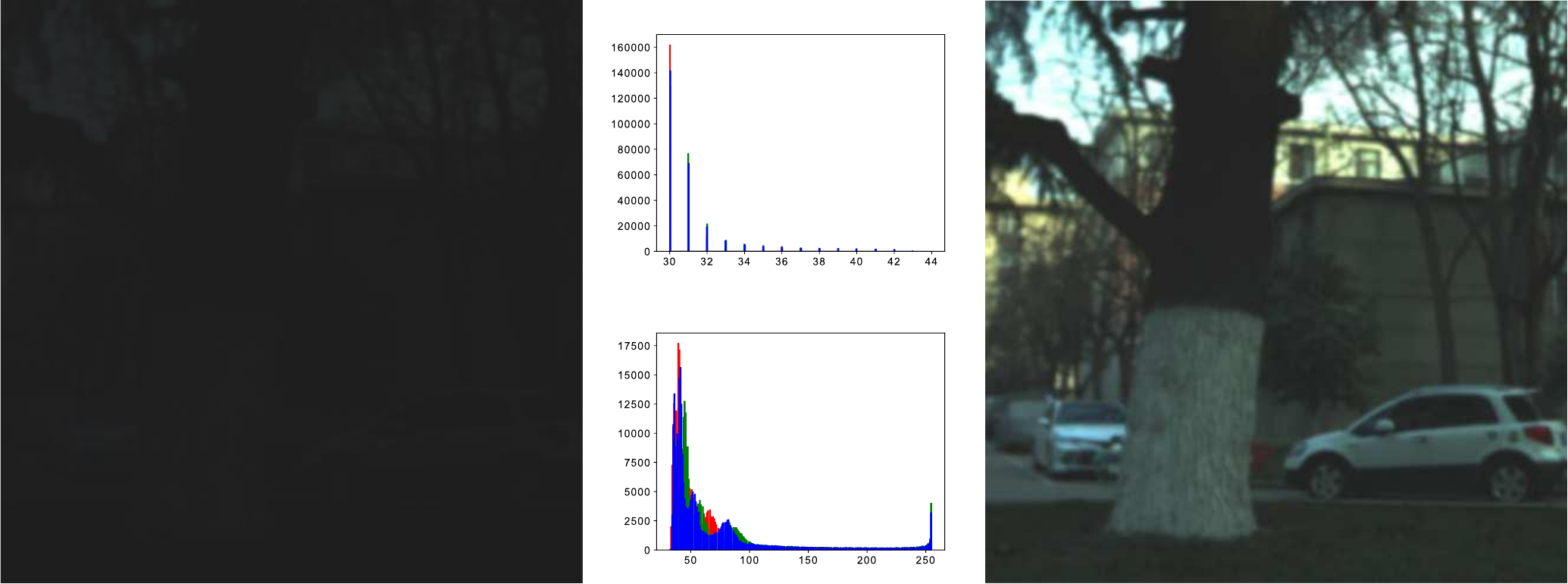}}
		\centerline{(c) Low-Frequency Component, Level=2, MSE=105.4 }\medskip
	\end{minipage}
	
	\vspace{-0.3cm}
	\caption
	{ (a) shows the original low-light and normal-light HSIs (pseudo-color with bands (57, 27, 17) captured with different light conditions. (b) is the Laplacian pyramids of the paired HSIs. As shown by the histograms, the differences between the low-light and normal-light HSIs are governed by the low-frequency components (c). Best viewed in color and zoomed in.}\medskip
	\label{motivation}  
	\vspace{-0.2cm}
\end{figure} 

The most straightforward strategy for enhancing low-light HSIs is to process the HSI through band scanning with a natural image enhancement method. In line with this strategy, traditional model-driven methods such as methods using statistical characteristic\cite{ibrahim2007brightness}, \cite{abdullah2007dynamic} and methods with Retinex theory\cite{wang2013naturalness}, \cite{fu2015probabilistic} can be directly employed for low-light HSI enhancement. Apart from traditional model-driven methods, deep-learning-based methods, such as SID\cite{sidchen2018learning}, Retinex-Net\cite{retinexnet_wei2018deep}, EnlightenGAN\cite{enlightengan}, and DRBN\cite{yang2020fidelity} can also be utilized to solve this problem. However, without considering the correlation between different HSI bands and noise suppression, this strategy usually results in spectral distortion and amplified noises. This statement can be verified by the comparison results in Section \ref{sec:experimental_on_indoor_dataset}, where the low-light HSI enhancement results with many methods of this strategy are reported. Another strategy is adopting well-studied HSI denoising algorithms such as BM4D\cite{bm4d}, total variation-based (TV) methods (LRTV) \cite{LRTV} and low rank methods \cite{LRMR}, \cite{he2019non}. However, these model-driven HSI denoising methods concentrate on denoising only and ignore promoting HSI contrast and visibility. Most recently, the success of deep learning also stimulates the development of HSI denoising \cite{hsid}, \cite{encam}, \cite{yuan2021partial}, and \cite{9732909}. Existing deep-learning-based methods usually perform well on HSI denoising tasks. However, they are not perfectly suitable to enhance low-light HSI, since they are not considering the intrinsic properties of low-light HSI. Therefore, how to leverage the intrinsic properties of low-light HSI is a fundamental problem for enhancement tasks.

The intrinsic properties of a low-light HSI are required to be exploited for designing an appropriate enhancement approach. Here we discover two important intrinsic properties of a low-light HSI according to its statistical characteristics. On one hand, in a low-light HSI, domain-specific attributes, such as illumination, are mainly related to low-frequency components, while the details of textures are relevant to high-frequency components\cite{liang2021high}. As depicted in Fig.~\ref{motivation}, we capture paired HSIs in the same scene with short and long exposure times, respectively. In Fig.~\ref{motivation}, the left column displays the HSIs captured in low-light condition, while the right column shows their counterpart in the normal-light condition. The mean squared errors between the low-frequency components (see Fig.~\ref{motivation} (c)) under the two conditions are about five times greater than those between the high-frequency (see Fig.~\ref{motivation} (b)) components. Paying attention to the histograms and pseudo-color visual appearance of the paired HSIs, a similar conclusion can be drawn. On the other hand, since information among different bands of the HSI is redundant and complementary, we conclude the averaged high-frequency component of adjacent bands in low-light HSI contains more textural information than that of a single band. Hence, we can restore the lost textures with low luminance values by averaging high-frequency components, leaving only textures with high luminance values to recover. This inspires us that the averaged high-frequency components of adjacent bands in low-light HSI is stronger prior than the high-frequency component of a single band.

Based on the findings, we propose the low-light HSI enhancement (HSIE) model. This model can first enlighten the dark areas of a low-light HSI. Then, it is capable of suppressing various noises and keeping spectral fidelity at the same time. In specific, we build a two-branch network, of which the overall structure is depicted in Fig.~\ref{fig:LHSIE}. The illumination enhancement branch aims to enlighten the low-frequency component of a low-light HSI. The high-frequency refinement branch intends to restore the textural details. The illumination enhancement branch consists of three sub-modules. The first sub-module is responsible for extracting multi-scale spatial and spectrum features of HSI. Then, the second sub-module is used to enlighten the dark areas and remove various noises in low-light HSI. Finally, the purpose of the third sub-module is to reconstruct the low-frequency component of a low-light HSI. We restore the high-frequency components of the low-light HSI through the high-frequency refinement branch. For the sake of efficiency, the high-frequency refinement branch is a lightweight network. It is composed of three cascaded residual blocks to predict a mask with which the textual details of HSI can be adaptively adjusted. In addition, the input of the high-frequency refinement branch is set as the average of the high-frequency components of adjacent bands instead of a single band. This design aims to sufficiently leverage the complementary properties among adjacent bands of HSI and ease the learning of mapping between high-frequency components.

Below is an overview of our major contributions.

\begin{enumerate}
\item We present a two-branch low-light Hyperspectral Images Enhancement network (HSIE) based on Laplacian pyramid decomposition and reconstruction, with two intrinsic properties of low-light HSI taken into account.  HSIE can effectively boost the brightness of low-light HSI, suppress noises, and efficiently keep spectral fidelity.
	
\item We have gathered a new LHSI dataset containing both indoor and outdoor scenes. To our knowledge, the LHSI dataset is the first to allow for the training and testing of low-light HSI enhancement approaches.

\item Promising results are achieved on the new LHSI dataset, which confirms the effectiveness of the proposed HSIE. Furthermore, some HSI classification experiments on the Indian Pine dataset are conducted to show that low-light HSI preprocessed by the proposed approach benefits the downstream tasks. 
\end{enumerate}

The remainder of this work is arranged in the following manner. In Section \ref{sec:rw}, we overview three related research areas, such as low-light natural image enhancement, HSI denoising, and the application of the Laplacian pyramid in deep-learning-based algorithms. The proposed HSIE is discussed in depth in Section \ref{sec:proposed_method}. In Section \ref{sec:experimental_analysis}, the experimental results and accompanying analysis on LSHI datasets are provided. In addition, the experimental results of a downstream classification task on the Indian Pine dataset and a denoising task on the Washington DC Mall dataset are also illustrated in Section \ref{sec:experimental_analysis}. Lastly, in Section \ref{sec:conclusion}, we conclude this work.

\section{Related Work} \label{sec:rw}

\subsection{Low-Light Natural Image Enhancement}

In recent years, low-light natural image enhancement has achieved significant progress. There are essentially two types of solutions to this problem, traditional model-driven methods, and deep-learning-based data-driven methods. Traditional model-driven methods are mainly based on statistical characteristics like \cite{ibrahim2007brightness}\cite{abdullah2007dynamic} and Retinex theory\cite{wang2013naturalness}\cite{fu2015probabilistic}, \cite{guo2016lime}\cite{park2017low}. In Retinex-based methods, they first obtain an illumination component as well as a reflectance component through decomposing the observed low-light natural image. Then, according to the Retinex theory\cite{land1977retinex}, the reflectance component holds steady under low-light scenes. Thus, the estimation of the illumination component dominates the enhancement results. Further, in \cite{wang2014variational}, the authors propose a classical variational Bayesian Retinex (VBR) method, which provides a new framework for Retinex from the perspective of Bayesian. However, most of the traditional methods only tackle the low-visibility problem without considering noise suppression, which leads to the amplification of noise in the enhanced results. Due to impressive performance gains and robustness over traditional methods, deep-learning-based low-light natural image enhancement approaches have attracted continuous attention\cite{li2021lighting}. Lore \emph{et al}. \cite{lore2017llnet} proposed LLNet made up of a deep autoencoder to simultaneously enlighten and denoise low-light natural images. Thereafter, a U-Shaped convolutional network is introduced by Chen \emph{et al}. \cite{sidchen2018learning} to map the short-exposure image with raw format to its full-exposed counterpart with sRGB format. In addition, Chen established a low-light dataset named SID. An unsupervised pioneering work EnligthenGAN\cite{enlightengan} was developed to eliminate the dependency on paired data. The generator of EnlightenGAN was based on an attention-guided U-Net, while the discriminator takes both global and local information into account\cite{zhao2021_rsgn9399800} to guarantee the reality of the enhanced image. Recently, a lightweight yet effective low-light enhancement method RUAS was proposed by Liu \emph{et al}. \cite{liu2021retinex}. By considering the Retinex rule, RUAS introduced a label-free learning strategy to exploit low-light prior characteristics for the illumination map in a carefully designed compact search space. Above all, the low-light natural image enhancement methods only consider three channels (R, G, and B) and cannot guarantee the spectral consistency between bands in HSI. This paper focuses on designing a low-light HSI enhancement model, which intends to restore all bands of an HSI.

\subsection{HSI Denoising}
HSI denoising aims to distill clean data from its noisy counterpart\cite{li2021videodistillation}. HSI denoising methods can be mainly categorized into two types, traditional model-driven, and deep-learning-based data-driven. BM4D\cite{bm4d}, an extended version of the well-known image denoising method BM3D\cite{dabov2007image}, is a classical traditional model-driven method for HSI denoising. However, BM4D suffers from an unacceptable long processing time for HSIs with large spatial resolution. Yuan \emph{et al}. \cite{hsid} proposed a data-driven CNN model with residual learning (HSID-CNN). The HSID-CNN can simultaneously extract multi-scale spatial and spectral joint features and achieve excellent performance. However, when the noise distribution is irregular among HSI bands, HSID-CNN fails to suppress noise efficiently due to a lack of feature representation. To achieve better denoising performance, Ma \emph{et al}. \cite{encam} proposed an attention-based\cite{zhao2018hsa} enhanced non-local cascading network. Another shortcoming of HSID-CNN is that it needs to train different models for varying noise levels. To overcome this limitation, Yuan \emph{et al}. \cite{liu2021retinex} introduced Partial-DNet with strong generalization ability, which estimates the noise map of every band of HSI as guidance for adaptively fitting different datasets. These methods are usually trained with small patches ($20 \times 20$) for the sake of training cost, thus these methods always have a limited receptive field, leading to performing well only on fixed distribution noises instead of inconsistent noises. To exploit the underlying global and local spatial-spectral characteristics of HSI, Wei \emph{et al}. \cite{wei20203} proposed QRNN3D, where RNN-based attention\cite{zhao2019cam} were employed. The architecture of QRNN3D is a typical encoder-decoder model with a residual connection. Recently, Shi \emph{et al}. \cite{3d_danet} proposed 3D-ADNet, which is an advanced version of HSID-CNN. To fully exploit the global correlation between spectral and spatial, 3D-ADNet employed the self-attention mechanism and a multiscale structure. Although all of these methods can be immediately expanded to enhance low-light HSI, neither one of them considers the intrinsic properties of low-light HSI. Besides, they are designed for different purposes, and the HSI denoising model is mainly for denoising but not for enhancing the brightness of the HSI. Thus, the performance of these methods is limited when applied to enhance the low-light HSI. We intend to specially design an algorithm for enhancing low-light HSI by taking the intrinsic properties of low-light HSI into account.

\subsection{Laplacian Pyramid}
The classical hierarchical structure of the Laplacian pyramid benefits several deep-learning-based tasks such as high-resolution image translation\cite{liang2021high}, unsupervised image generation\cite{denton2015deep}, and image super resolution\cite{lai2017deep}. Denton \emph{et al}. \cite{denton2015deep} proposed Laplacian Generative Adversarial Networks, which aim to generate sharp images by training Generative Adversarial Networks at each level of the Laplacian pyramid. Lai \emph{et al}. \cite{lai2017deep} first generates multiple high-frequency residuals serving as different levels of the Laplacian pyramid, then reconstructs the final image gradually using the Laplacian pyramid. To alleviate the heavy computation burden when translating high-resolution images, Liang \emph{et al}. \cite{liang2021high} designed Laplacian Pyramid Translation Network (LPTN). LPTN is composed of two main branches, of which one is used to translate the low-frequency component of an image with reduced resolution, while the other aims to refine the high-frequency component through a progressive masking strategy. The proposed HSIE differs from LPTN in that LPTN intends to translate natural RGB images while HSIE concentrates on enhancing low-light HSIs. 

\section{PROPOSED APPROACH} \label{sec:proposed_method} 

\subsection{Overview} \label{model_overview}

\begin{figure*}[!htb]
	
	\begin{minipage}[b]{1.0\linewidth}
		\centering
		{\includegraphics[width= 1.0\textwidth]{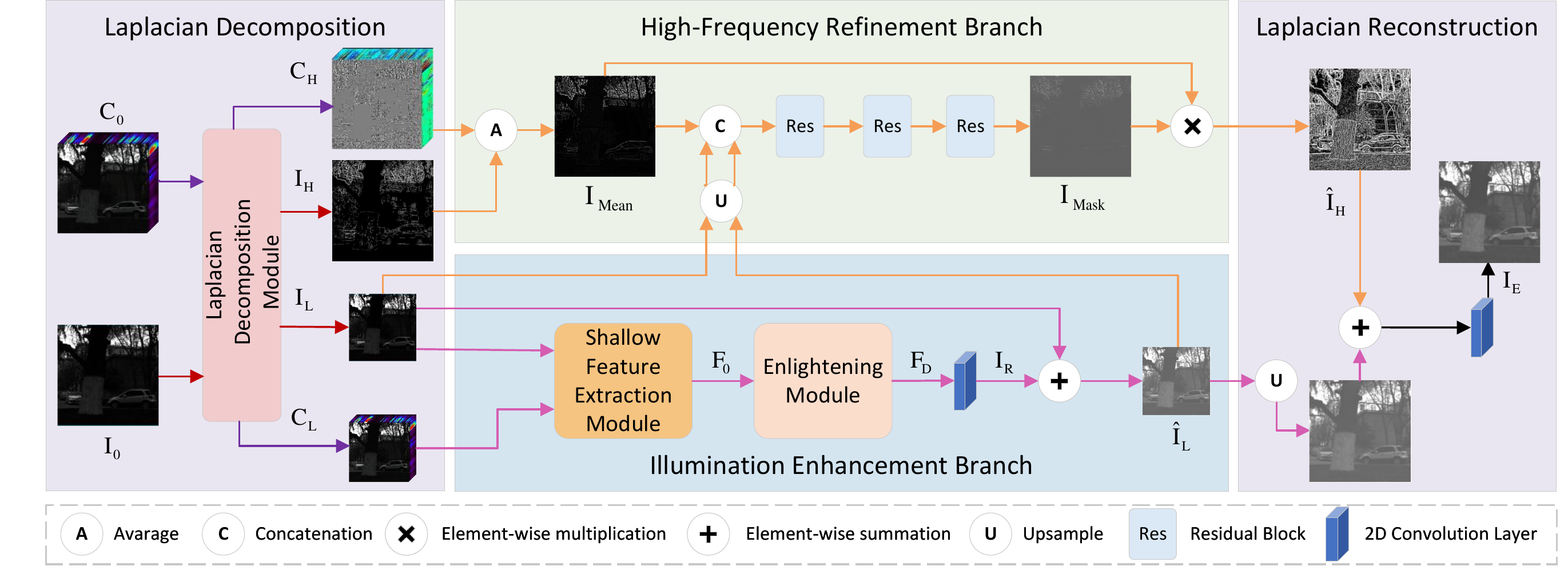}}
		
	\end{minipage}
	\caption
	{ The overall structure of the proposed HSIE. Given a band of HSI $I_{0}\in\mathbb{R}^{h\times w\times 1}$ and its adjacent $k$ bands $C_{0}\in\mathbb{R}^{h\times w\times k}$, we first decompose $I_{0}$ and $C_{0}$ into an Laplacian pyramid respectively. The Laplacian Decomposition Module represents the standard Laplacian decomposition process. \textcolor[rgb]{0.44,0.19,0.63}{Purple} arrows: We decompose $C_{0}$ into $C_{H}$ and $C_{L}$. \textcolor[rgb]{0.75,0,0}{Red} arrows: We decompose $I_{0}$ into $I_{H}$ and $I_{L}$. \textcolor[rgb]{0.84,0.36,0.69}{Pink} arrows: The low-frequency component $I_L\in\mathbb{R}^{\frac{h}{2}\times \frac{w}{2}\times 1}$ is enhanced using a network with shallow feature extraction module (SFE) and enlightening module (EM). \textcolor[rgb]{0.96,0.64,0.39}{Brown} arrows: The $C_{H}\in\mathbb{R}^{h\times w\times k}$ together with $I_{H}$ are averaged to get an informative high-frequency prior $I_{Mean}\in\mathbb{R}^{h\times w\times 1}$. To refine the high-frequency component $I_{Mean}$, we learn a mask $I_{Mask}\in\mathbb{R}^{h\times w\times 1}$ with lightweight residual blocks. }\medskip
	\label{fig:LHSIE} 
	\vspace{-0.6cm}
\end{figure*}

To overcome the low-light HSI enhancement challenge, we present an end-to-end model called low-light hyperspectral image enhancement network (HSIE). Fig.~\ref{fig:LHSIE} depicts the overall structure of the proposed HSIE. The output of HSIE can be formulated as
\begin{align}
	\begin{split}
		I_{E}=H_{HSIE}\left ( I_{0}, C_{0} \right ),
	\end{split}
\end{align}  
where $H_{HSIE}$ denotes the function of the proposed HSIE. $I_{0}\in\mathbb{R}^{h\times w\times 1}$,  $C_{0}\in\mathbb{R}^{h\times w\times k}$ and  $I_{E}\in\mathbb{R}^{h\times w\times 1}$ denote a low-light band, its adjacent $k$ bands, and the output enhanced band, respectively. Among the adjacent $k$ bands of the data cube $C_{0}$, the first half is sampled before band $I_{0}$, while the other half is sampled after band $I_{0}$. We can get the full enhanced HSI through iterating over each band of the low-light HSI.

As depicted in Fig.~\ref{fig:LHSIE}, we first decompose $I_{0}$ into a Laplacian pyramid through the Laplacian Decomposition Module, obtaining a high-frequency component represented by $I_{H}$ and a low-frequency component $I_{L}$. The resolution of $I_{H}$ is $h\times w$, while the width and height of $I_{L}$ is $\frac{h}{2}$ and $\frac{w}{2}$ respectively. The Laplacian Decomposition Module stands for a standard Laplacian decomposition process. Due to the invertible characteristic of the Laplacian pyramid, we can reconstruct the original image by sequenced mirror operations. According to Burt and Adelson~\cite{burt1983laplacian}, $I_L$ is blurred by a Gaussian filter and reflects the global attributes of an image. Concurrently, $I_{H}$ demonstrates detailed textures of the image where most pixels have an intensity value close to $0$. At the same time, we carry out this decomposition operation on data cube $C_{0}$. As a result, we obtain $C_{H}$ and $C_{L}$, which denote the resolution-kept high-frequency component and the resolution-reduced low-frequency component of the data cube, respectively. To leverage the low-rank properties of HSI\cite{he2019non}, we start from the averaged $I_{Mean}$ of $I_{H}$ and $C_{H}$ instead of $I_{H}$ to refine the high-frequency component, since $I_{Mean}$ involves parts of texture information lost in $I_{H}$. 

Inspired by the above properties of the Laplacian pyramid and low-light HSI, we propose to enhance $I_L$ to recover the illumination, meanwhile, refine $I_{Mean}$ adaptively to reduce artifacts in reconstruction. The proposed HSIE model is therefore composed of two branches. For the first branch, we convert the low-resolution $I_{L}$ to $\hat{I}_{L}$ using a convolution network comprised of three modules, a shallow feature extraction module, an enlightening module, and a reconstruction module. Residual learning strategy\cite{DnCNN}, which was tested to be effective for restoration tasks, is applied in this branch to stabilize and speed up the convergence process. For the second branch, we learn a mask of the high-frequency component through a lightweight convolution network with the concatenation of $[I_{Mean}, up(I_{L}), up(\hat{I}_{L})]$ as input, where $up(\cdot)$ represents the bilinear upsampling operation. To refine the high-frequency component $I_{Mean}$, the mask is applied to $I_{Mean}$ by pixel-wise multiplication. We introduce the two branches in detail in the following sections.

\subsection{Illumination Enhancement Branch}

\subsubsection{Shallow Feature Extraction}
\label{ssec:subhead}

\begin{figure}[t]
	\centering
	\includegraphics[width=0.98\columnwidth]{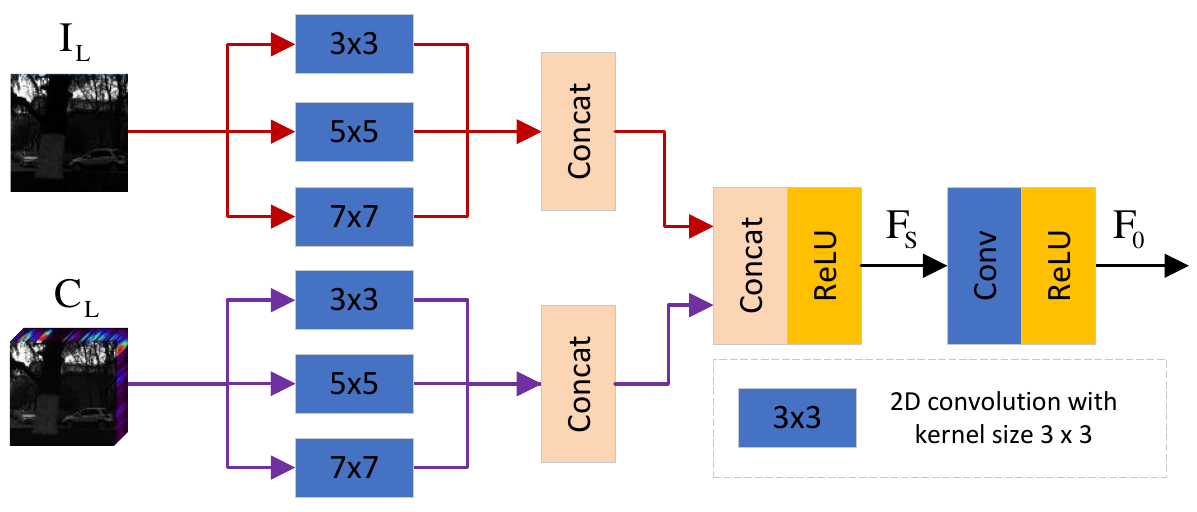}
	\caption{Shallow Feature Extraction Module.}
	\label{fig:shallow_feature_extraction_module}
\end{figure}

The shallow feature extraction module is composed of two submodules as shown in Fig.~\ref{fig:shallow_feature_extraction_module}. The first submodule takes a low-frequency component $I_{L}$ of the low-light band $I_{0}$ as its input and is composed of three different kernel sizes for 2D convolution to extract multi-scale spatial information. The three convolutions are executed concurrently and the extracted features are concatenated to form a feature map $F_{SL}$. This process can be represented as
\begin{align}
	\begin{split}
		F_{SL}&=\left [ H_{L3} (I_{L}),H_{L5} (I_{L}), H_{L7} (I_{L}) \right ],
	\end{split}
\end{align}  
where $ H_{L3}$, $ H_{L5}$ and $ H_{L7}$ denote 2D convolution with kernel size $3 \times 3$, $5 \times 5$, and $7 \times 7$, respectively. The second submodule takes a low-frequency component $C_{L}$ of the low-light data cube $C_{0}$ as input. This is motivated by the fact that the redundant spectral information in an HSI is beneficial for restoring hyperspectral image\cite{hsid}. The second submodule extracts multi-scale joint spatial-spectral representation and shares the same configuration as the first submodule. The extracted representation is then concatenated to form a representation $F_{SC}$, the same size as the output of the first submodule. This process can be represented as
\begin{align}
	\begin{split}
		F_{SC}&=\left [ H_{C3} (C_{L}), H_{C5} (C_{L}), H_{C7} (C_{L}) \right ],
	\end{split}
\end{align}  
where $ H_{C3}$, $ H_{C5}$ and $ H_{C7}$ denote 2D convolution with different kernel sizes executed on the low-frequency component $C_{L}$ of the low-light data cube $C_{0}$.

Finally, $F_{S}$ is obtained by concatenating the output of the two submodules. This process can be represented as
\begin{align}
	\begin{split}
		F_{S}&=ReLU\left ( \left [  F_{SL}, F_{SC} \right ] \right ),
	\end{split}
\end{align}  
where $ReLU$ denotes the rectified linear units (ReLU) function.

The output feature map $F_{S}$ is introduced directly to a convolution layer to half the channels and further extracts features. This operation can be denoted as 
\begin{align}
	\begin{split}
		F_{0}&=ReLU\left (H_{Conv}\left ( F_{S} \right ) \right ),
	\end{split}
\end{align}  
where $H_{Conv}$ denotes a $3 \times 3$ convolution layer with padding 1 to maintain the feature map's spatial resolution. The feature map $F_{0}$ extracted by this operation is set as the input to the Enlightening Module. 

\subsubsection{Enlightening Module}
\label{ssec:subhead}

The enlightening module is mainly composed of several CABs, as shown in Fig.~\ref{fig:RDECAB} (top). Supposing we have $N$ CABs, the output $F_{i}$ of the $i$-th CAB can be obtained by 
\begin{align}
	\begin{split}
		F_{i}&=H_{CAB,i}\left ( F_{i-1} \right )\\
		&=H_{CAB,i}\left ( H_{CAB,{i-1}}\left ( \cdots \left ( H_{CAB,1}\left ( F_{0} \right ) \right ) \cdots \right ) \right ),
	\end{split}
\end{align} 
where $H_{CAB, i}$ represents the operation of the $i$-th CAB. $H_{CAB, i}$ is composed of several standard operations, namely 2D convolution, global average pooling, and activation function. The details of CAB will be discussed in the following sections. Following the extraction of deep features with a list of CABs, a feature fusion function is applied to integrate the output features from all of the previous CABs. This process can be denoted as
\begin{align}
	\begin{split}
		F_{D}=H_{DF}\left ( \left [ F_{0},F_{1},\cdots ,F_{N} \right ] \right ),
	\end{split}
\end{align}    
where $\left [ F_{0},F_{1},\cdots ,F_{N} \right ]$ denotes the concatenation of feature-maps. $H_{DF}$ is a standard convolution with kernel size $1\times 1$, which is leveraged to integrate features from different levels.

\subsubsection{Residual Dense Effective Channel Attention Blocks} 

\begin{figure}[t]
	\centering
	\includegraphics[width=0.98\columnwidth]{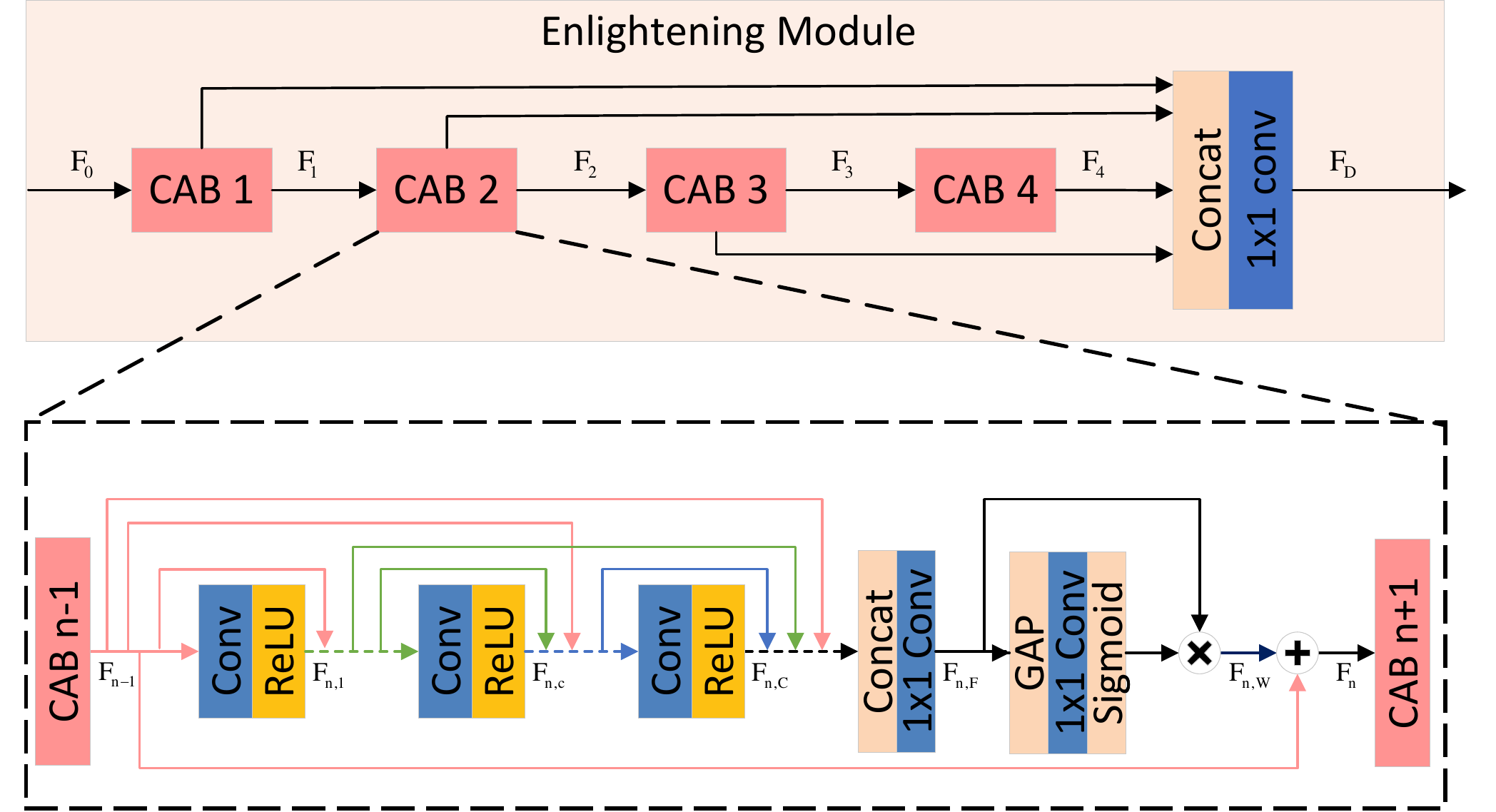}
	\caption{The enlightening module (top) and the residual dense effective channel attention block (CAB) (bottom).}
	\label{fig:RDECAB}
\end{figure}

The detailed structure of a CAB is depicted in Fig.~\ref{fig:RDECAB} (bottom). The CAB consists of several dense connected layers\cite{densenet_huang2017densely}, an effective channel attention\cite{wang2020eca} module and a residual connection. In Fig.~\ref{fig:RDECAB} (bottom), $F_{n-1}$ and $F_{n}$ symbolize the input and output feature map of the $n$-th CAB, respectively. Feature map generated from the $c$-th convolution layer of the $n$-th CAB can be obtained by
\begin{align}
	\begin{split}
		\label{eq:F_n_c}
		F_{n,c}=ReLU \left ( W_{n,c}\left [ F_{n-1},F_{n,1},\cdots ,F_{n,c-1} \right ]  \right ),
	\end{split}
\end{align}
 where $W_{n,c}$ denotes the shared parameters of the $c$-th convolution layer of which the bias term is neglected to simplify the expression. The feature-maps generated by the $\left (n - 1 \right)$-th CAB were concatenated, which is denoted as $\left [ F_{n-1},F_{n,1},\cdots ,F_{n,c-1} \right ]$. There exist direct connections between the precedent CAB's output feature map and every layer of the current CAB. In addition, each layer in the same CAB has direct links to all succeeding layers. This type of dense connection can strengthen feature propagation and encourage feature extraction.

A in-block feature fusion operation is followed to merge all the feature-maps produced by each convolution layer and the output feature-map of the proceeding CAB, which can be formulated as
\begin{align}
	\begin{split}
		\label{eq:F_d_LF}
		F_{n,F}=H_{T}\left ( \left [ F_{n-1},F_{n,1},\cdots ,F_{n,c},\cdots ,F_{n,C}\right ] \right ),
	\end{split}
\end{align} 
where $H_{T}$ represents a transition layer composed of a single convolution layer with kernel size $1 \times 1$ , which is used to decrease the dimension of the concatenated feature maps. This transition layer alleviates the training and computational burden of the whole network and makes the network easy to train. 

After obtaining the fused feature-map, an effective channel attention module is adopted to capture inter-channel interaction efficiently. Assume we obtained a fused feature-map $F_{n, F} \in \mathbb{R}^{W \times H \times C}$, where $W$ and $H$ denotes feature-map's width and height respectively, and the channel dimension of $F_{n, F}$ is represented by $C$. The weights of the channels in the effective channel attention module can be obtained by
\begin{align}\label{fun_eca_weight}
	W_{A} = \sigma(H_{1D} (g(F_{n, F}))),
\end{align}
where $F_{n, F}$ is the output of the in-block feature fusion operation and $g(F_{n, F})=\frac{1}{WH}\sum_{i=1,j=1}^{W,H}F_{n, F}$ indicates global average pooling. The $H_{1D}$ denotes 1D convolution with kernel size $3$, which avoids dimensionality reduction and captures channel attention in an efficient way. The $\sigma$ is a Sigmoid function. We can obtain attention weighted feature-map by
\begin{align}\label{fun_f_n_W}
	F_{n, W} = F_{n, F} \otimes W_{A},
\end{align}
where $\otimes$ denotes element-wise multiplication.

To further strengthen representation ability of our model, the inner-block residual learning strategy is applied. The $n$-th CAB's final output may be written as
\begin{align}\label{fun_final_output_RDECAB}
	F_{n} = F_{n, W} + F_{n-1}.
\end{align}

\subsubsection{Reconstruction Module}

We reconstruct the enlightened low-frequency component of the HSI band by a single $3 \times 3$ convolution layer denoted as $H_{R}$, which is used to restore the residual $I_{R}$ directly (See Fig.~\ref{fig:LHSIE}). This procedure can be represented as
\begin{align}
	\begin{split}
		I_{R}=H_{R}\left (F_{D} \right ),
	\end{split}
\end{align} 
where $F_{D}$ denotes the Enlightening Module's output and $I_{R}$ represents the restored residual.

Finally, the restored low-frequency component is obtained by simply adding the restored residual $I_{R}$ and the input low-light band $I_{L}$, which is formulated as
\begin{align}
	\begin{split}
		\hat{I}_{L}=I_{R} + I_{L},
	\end{split}
\end{align} 
where $\hat{I}_{L}$ is the enlightened low-frequency component of an HSI band.

\subsection{High-Frequency Refinement Branch}

To reach a reliable reconstruction, we refine the averaged high-frequency component $I_{Mean}$ on the basis of  $I_{L}$ and $\hat{I}_{L}$. We intend to generate a single-channel mask for the averaged high-frequency component $I_{Mean}$ and adjust $I_{Mean}$ using the generated mask in this branch. 

To match the resolution of $I_{Mean}\in\mathbb{R}^{h\times w\times 1}$, we first upsample $I_{L}\in \mathbb{R}^{\frac{h}{2}\times \frac{w}{2}\times 1}$ and $\hat{I}_{L}\in \mathbb{R}^{\frac{h}{2}\times \frac{w}{2}\times 1}$. Then, we feed the concatenated $[I_{Mean}, I_{L}, \hat{I}_{L}]$ into a lightweight network composed of three residual blocks, where the detailed composition of the lightweight network is depicted in Fig.~\ref{fig:LHSIE}.  A mask of $I_{Mean}$ is generated by the lightweight network, which is denoted as $I_{Mask}\in\mathbb{R}^{h\times w \times 1}$. We regard the mask as a global adjustment to the high-frequency component which is easier to be optimized than the images without decomposition. Hence, we adjust the $I_{Mean}$ by:
\begin{equation}
	\label{refine}
	\hat{I}_{H} =  I_{Mean} \otimes I_{Mask},
\end{equation}
where $\otimes$ indicates the pixel-wise multiplication. We can then reconstruct the resulting image $ I_{E} $ using the refined $ \hat{I}_{H} $ and the enhanced $\hat{I}_{L}$ according to the Laplacian pyramid reconstruction strategy. This can be formulated as 
\begin{equation}
	\label{refine}
	I_{E} = H_{R}\left (\hat{I}_{H} + Upscale(\hat{I}_{L})\right ),
\end{equation}
where $Upscale$ is an upsampling process, in which we first double the size of $\hat{I}_{L}$ by padding zero, then we convolve the resized image with a gaussian kernel, which is the same kernel used in the Laplacian pyramid decomposition. $H_{R}$ is a simple same convolution layer to further refine the reconstructed band.

\section{EXPERIMENTAL ANALYSIS}  \label{sec:experimental_analysis}

\subsection{Datasets}

\begin{figure}[htb]
	
	\begin{minipage}[b]{0.32\linewidth}
		\centering
		\centerline{\includegraphics[width=2.7cm,height=2.7cm]{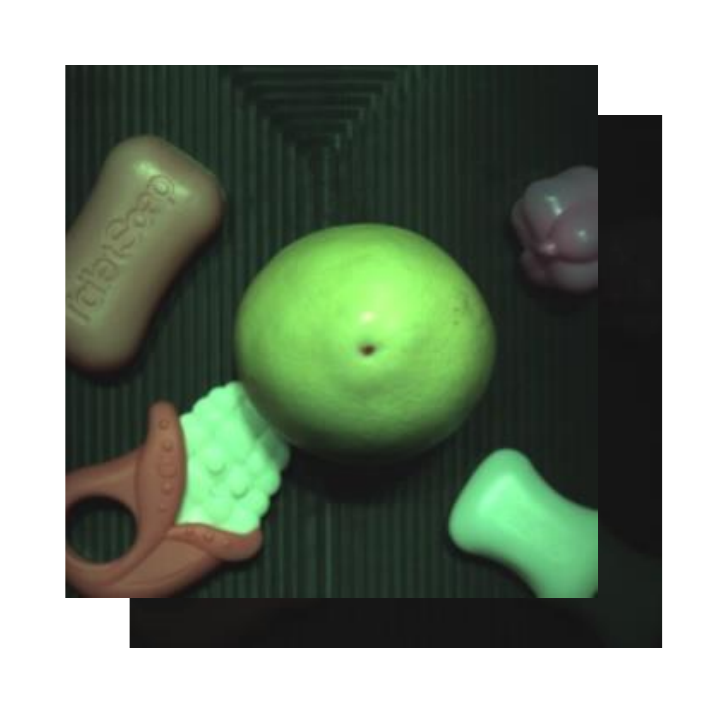}}
		\centerline{(a) Orange}\medskip
	\end{minipage}
	\hfill
	\begin{minipage}[b]{0.32\linewidth}
		\centering
		\centerline{\includegraphics[width=2.7cm,height=2.7cm]{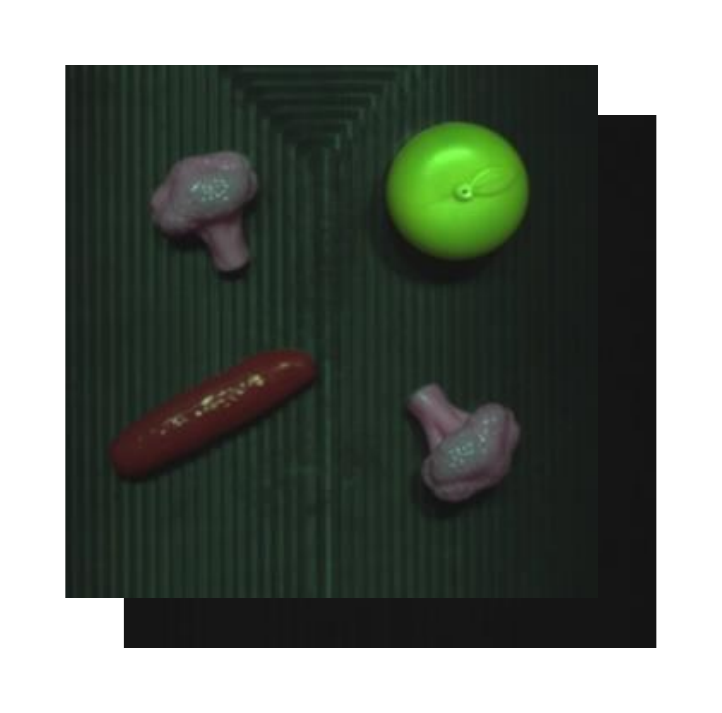}}
		\centerline{(b) Cauliflower}\medskip
	\end{minipage}
	\hfill
	\begin{minipage}[b]{0.32\linewidth}
		\centering
		\centerline{\includegraphics[width=2.7cm,height=2.7cm]{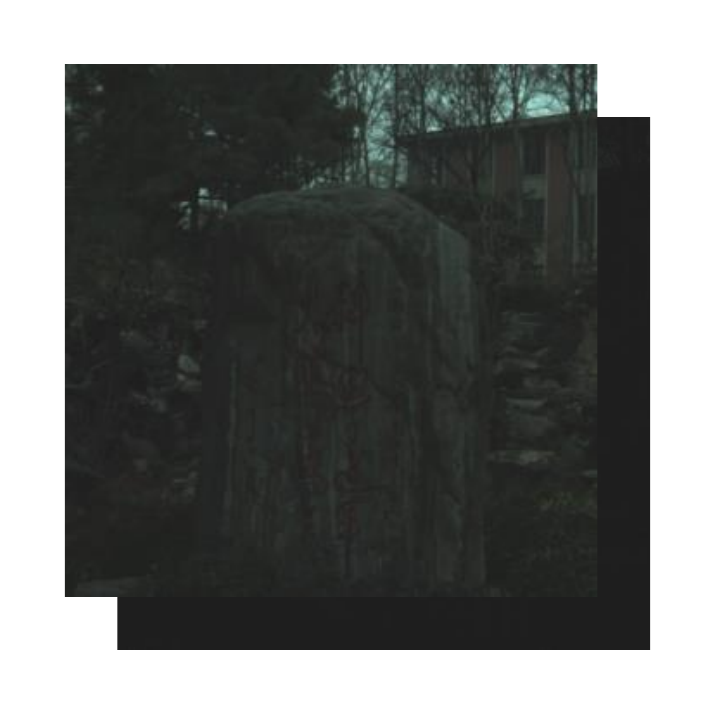}}
		\centerline{(c) Stone}\medskip
	\end{minipage}
	\begin{minipage}[b]{0.32\linewidth}
		\centering
		\centerline{\includegraphics[width=2.7cm,height=2.7cm]{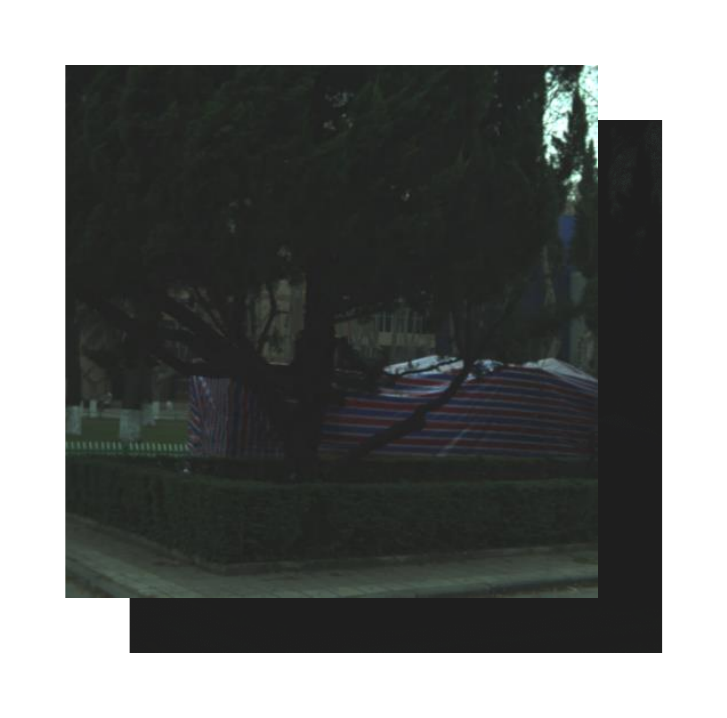}}
		\centerline{(d) Tree}\medskip
	\end{minipage}
	\hfill
	\begin{minipage}[b]{0.32\linewidth}
		\centering
		\centerline{\includegraphics[width=2.7cm,height=2.7cm]{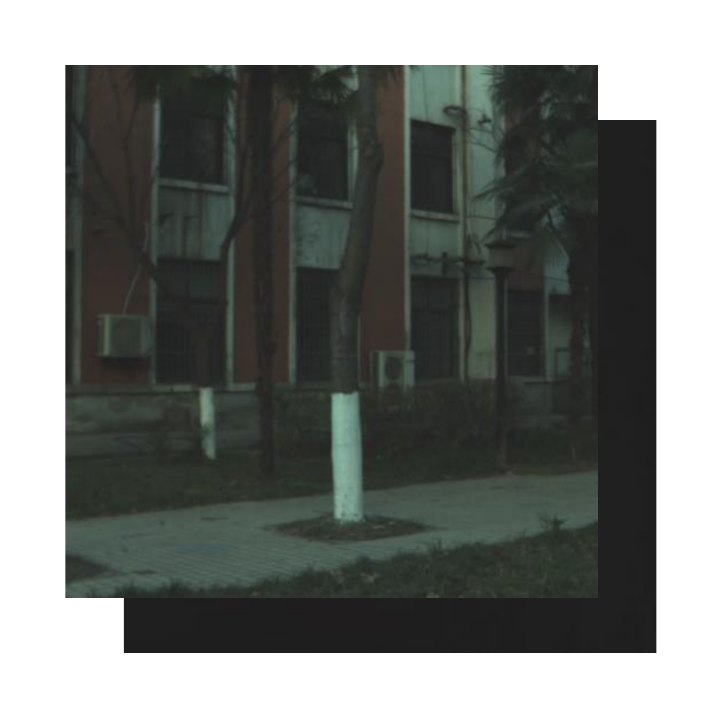}}
		\centerline{(e) Building}\medskip
	\end{minipage}
	\hfill
	\begin{minipage}[b]{0.32\linewidth}
		\centering
		\centerline{\includegraphics[width=2.7cm,height=2.7cm]{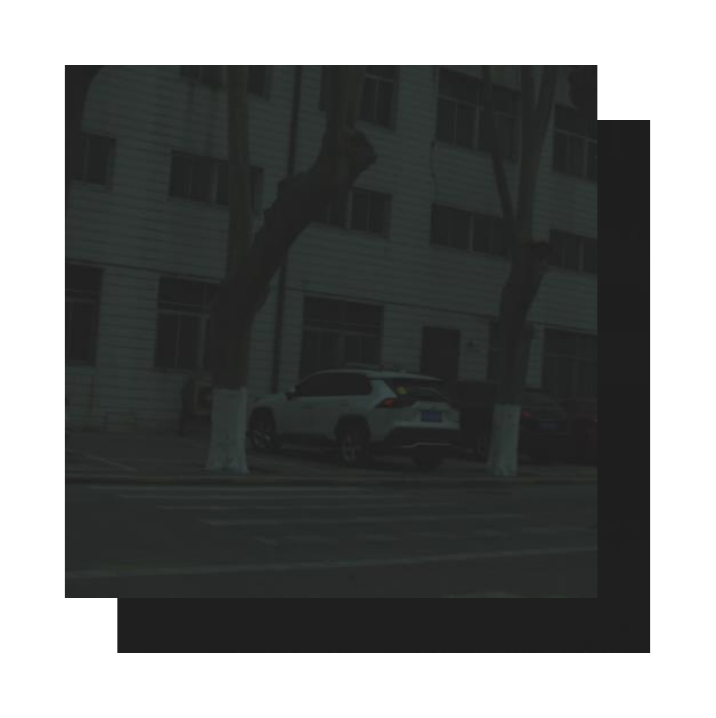}}
		\centerline{(f) Car}\medskip
	\end{minipage}
	
	\vspace{-0.3cm}
	\caption
	{ Pseudo-color example images in the LHSI dataset. (a) and (b) are indoor HSIs. (c) - (f) are outdoor HSIs. Pseudo-color Images of full-exposed HSIs (label) are displayed in the front. Pseudo-color images of short-exposure HSIs (basically dark) are shown in the back. Best viewed in color.}\medskip
	\label{fig:show_dataset_sample}  
	\vspace{-0.6cm}
\end{figure} 

A new low-light hyperspectral image (LHSI) dataset is gathered for the development of low-light HSI enhancement methods. Each sample in the LHSI dataset contains a short-exposure low-light HSI captured by setting the exposure time to 1ms and a long-exposure reference HSI captured with an exposure time of 15ms. The long-exposure HSI is sufficient to serve as ground truth. The LHSI dataset contains both indoor and outdoor sceneries. A small fraction of samples in the LHSI dataset is shown in Fig.~\ref{fig:show_dataset_sample}.

The indoor dataset is captured using the SPECIM FX10\footnote{Detailed information about the SPECIM FX10 can be found at \url{https://www.specim.fi/fx/}}, which works in a line-scan mode and is equipped with a fixed platform. During image acquisition in indoor scenes, we use a halogen light source to stabilize the lighting conditions. To make the HSI images acquired by the camera have accurate reflectance values, we first use a whiteboard for reflectance calibration before starting the acquisition. After calibration, the camera automatically converts the Digital Number (DN) values to reflectance values and outputs the reflectance spectrum image. The indoor dataset contains 6 pairs of HSIs, of which 5 pairs are used for training and 1 pair is used for testing. The spatial resolution of each indoor HSI is $390 \times512$, and each HSI has 224 bands with wavelengths ranging from 400nm to 1000nm. Since the first several bands and the last several bands are subjected to greater interference, we remove the first 20 bands and the last 12 bands. Among the remaining 192 bands, every three bands are equally spaced to choose one band, resulting in an HSI with the size of $390 \times512 \times64$, where $64$ represents the band number.

\begin{figure}[htb]
	
	\begin{minipage}[b]{0.32\linewidth}
		\centering
		\centerline{\includegraphics[width=2.7cm,height=2.7cm]{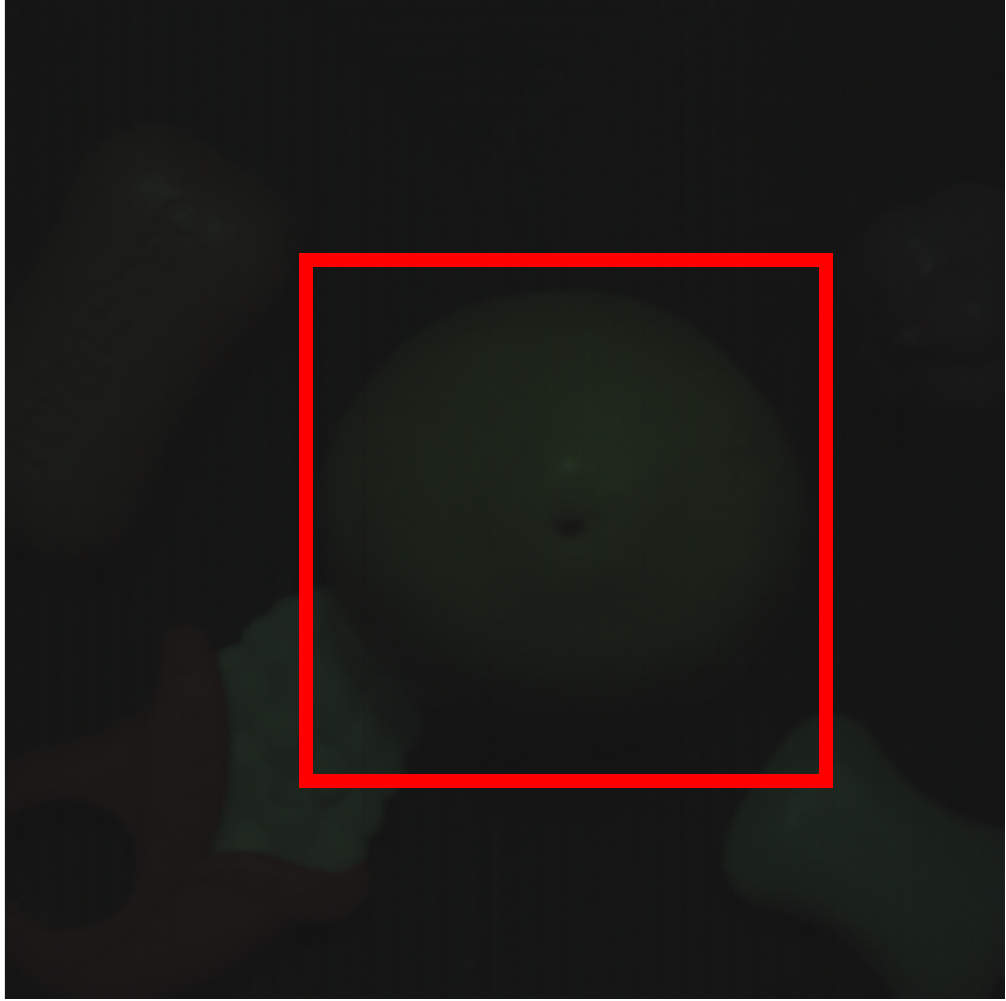}}
		\centerline{(a) Low-light}\medskip
	\end{minipage}
	\hfill
	\begin{minipage}[b]{0.32\linewidth}
		\centering
		\centerline{\includegraphics[width=2.7cm,height=2.7cm]{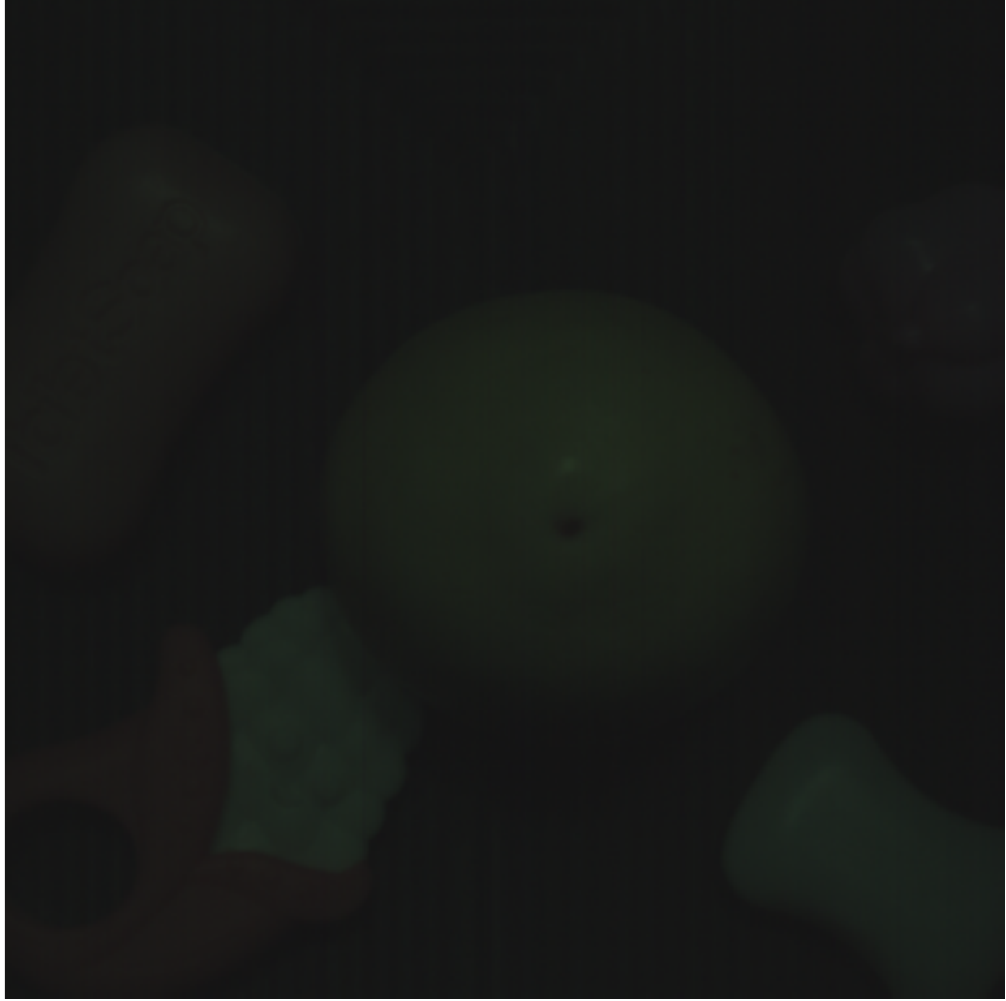}}
		\centerline{(b) MR}\medskip
	\end{minipage}
	\hfill
	\begin{minipage}[b]{0.32\linewidth}
		\centering
		\centerline{\includegraphics[width=2.7cm,height=2.7cm]{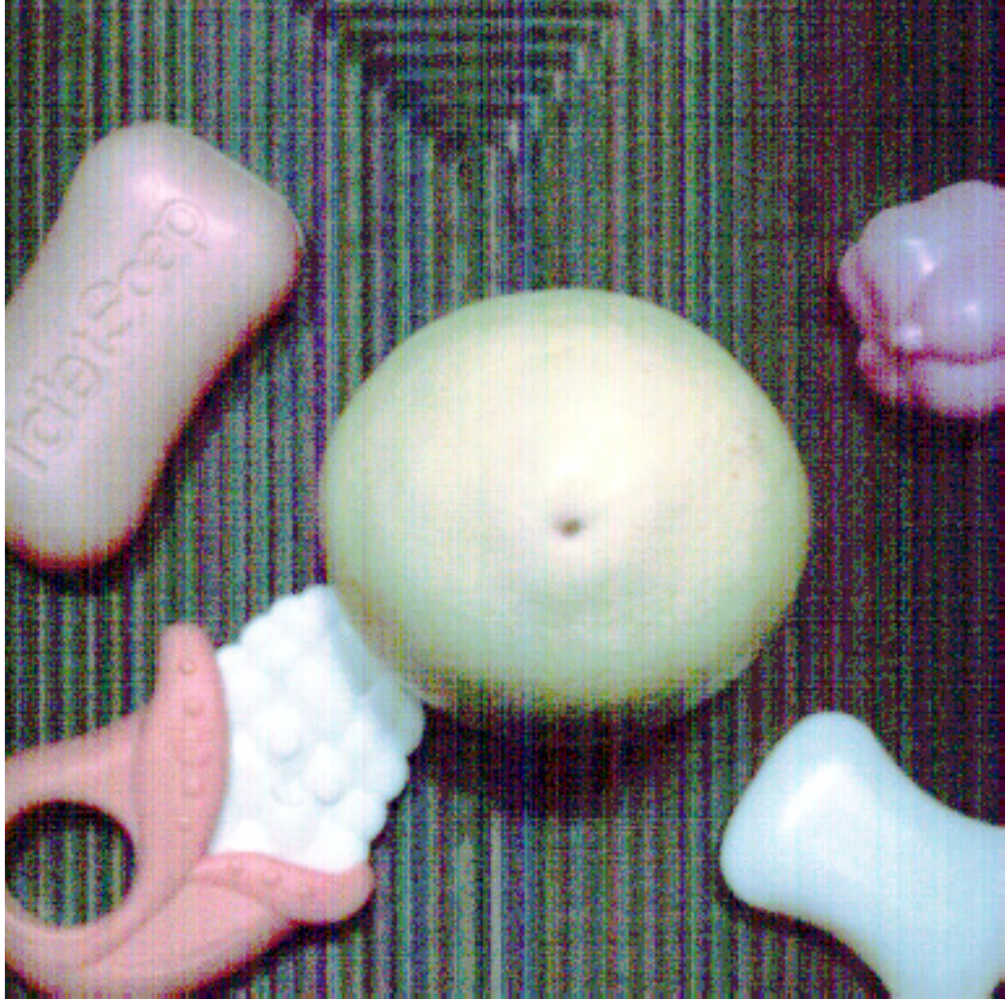}}
		\centerline{(c) HE}\medskip
	\end{minipage}
	\begin{minipage}[b]{0.32\linewidth}
		\centering
		\centerline{\includegraphics[width=2.7cm,height=2.7cm]{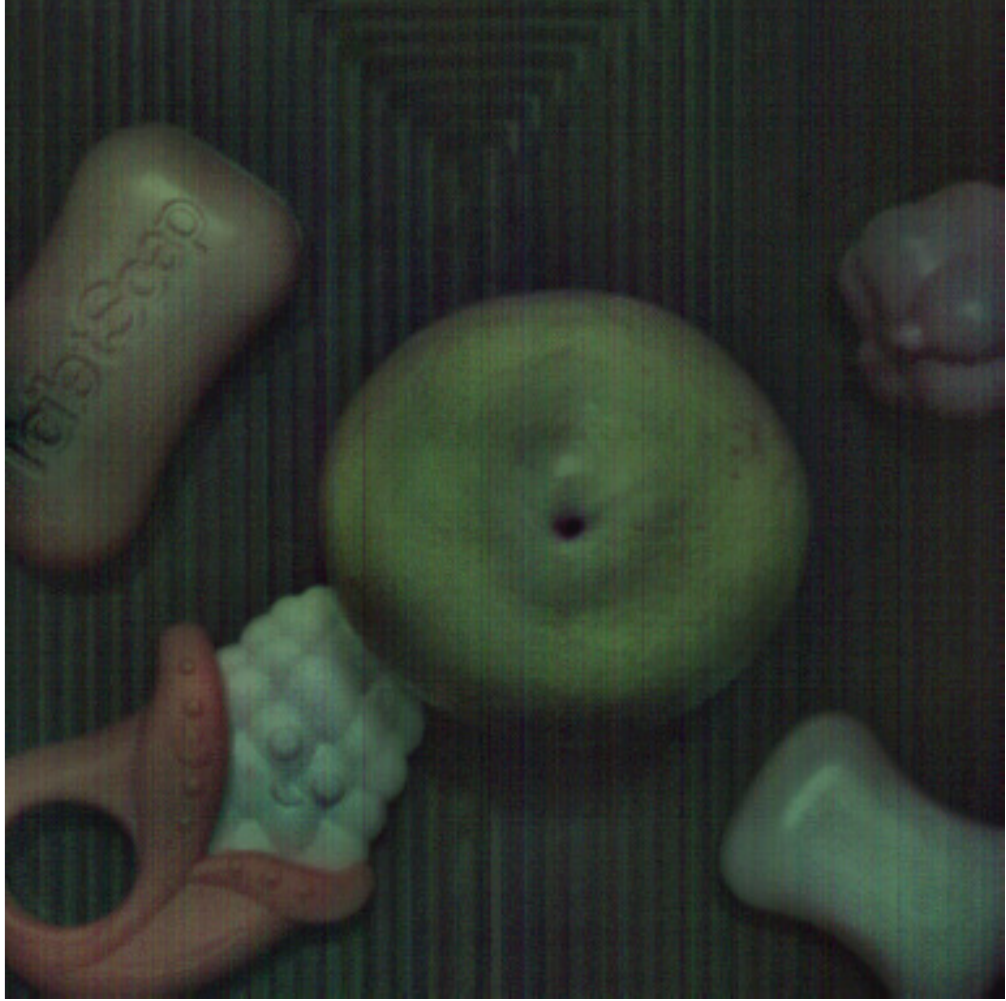}}
		\centerline{(d) CLAHE}\medskip
	\end{minipage}
	\hfill
	\begin{minipage}[b]{0.32\linewidth}
		\centering
		\centerline{\includegraphics[width=2.7cm,height=2.7cm]{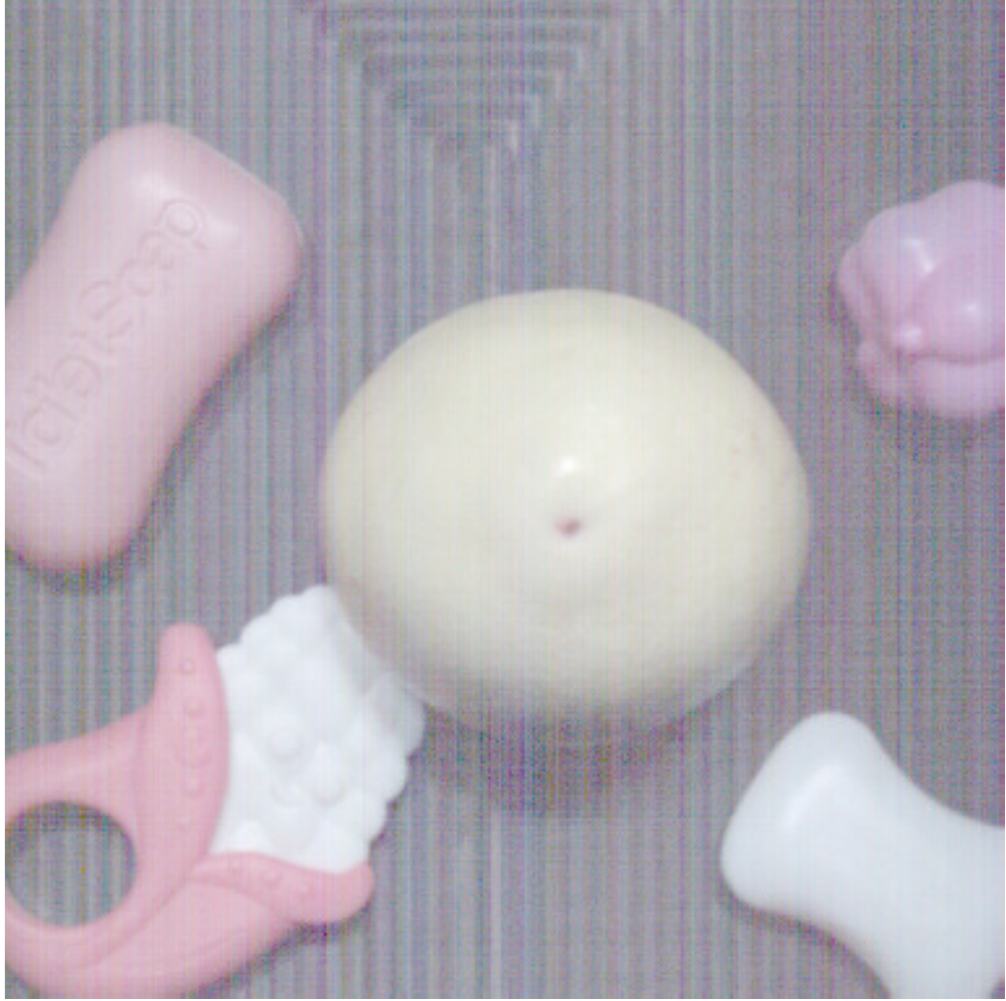}}
		\centerline{(e) MSR}\medskip
	\end{minipage}
	\hfill
	\begin{minipage}[b]{0.32\linewidth}
		\centering
		\centerline{\includegraphics[width=2.7cm,height=2.7cm]{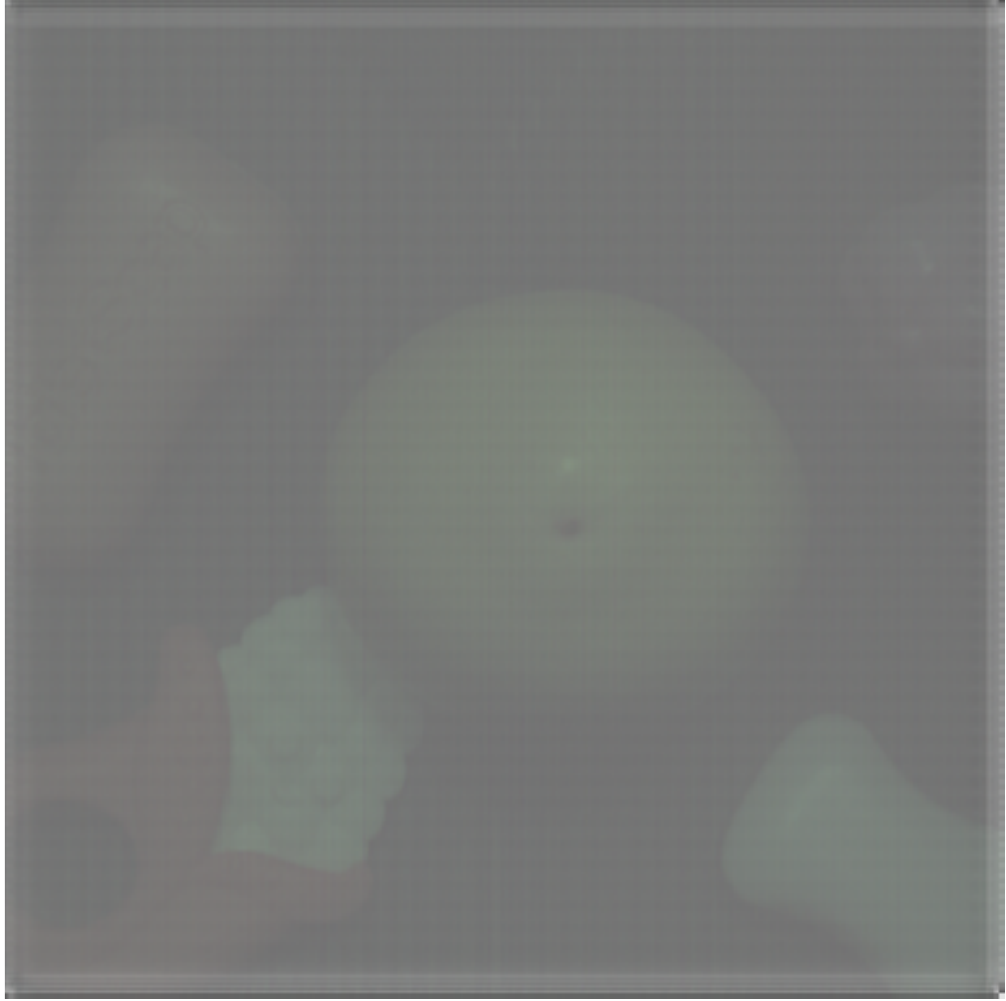}}
		\centerline{(f) Retinex-Net}\medskip
	\end{minipage}
	\begin{minipage}[b]{0.32\linewidth}
		\centering
		\centerline{\includegraphics[width=2.7cm,height=2.7cm]{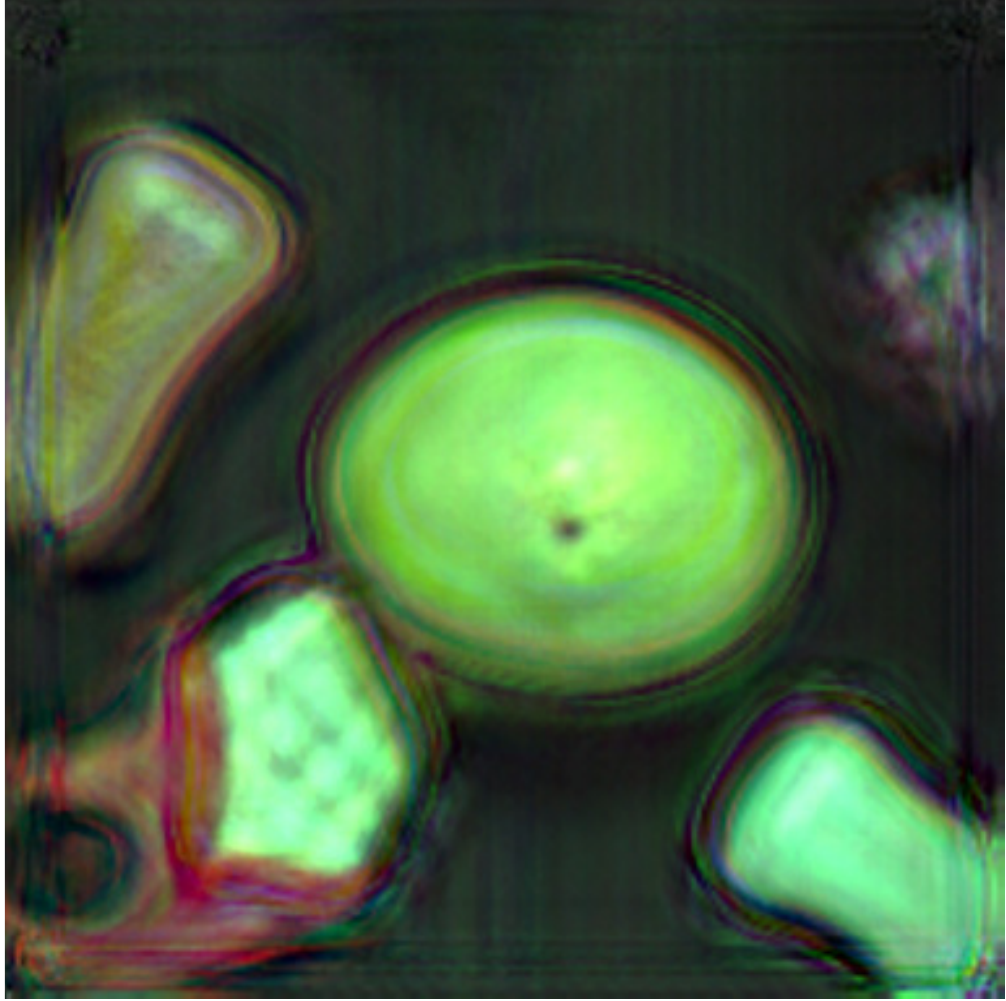}}
		\centerline{(g) 3D-ADNet}\medskip
	\end{minipage}	
	\hfill
	\begin{minipage}[b]{0.32\linewidth}
		\centering
		\centerline{\includegraphics[width=2.7cm,height=2.7cm]{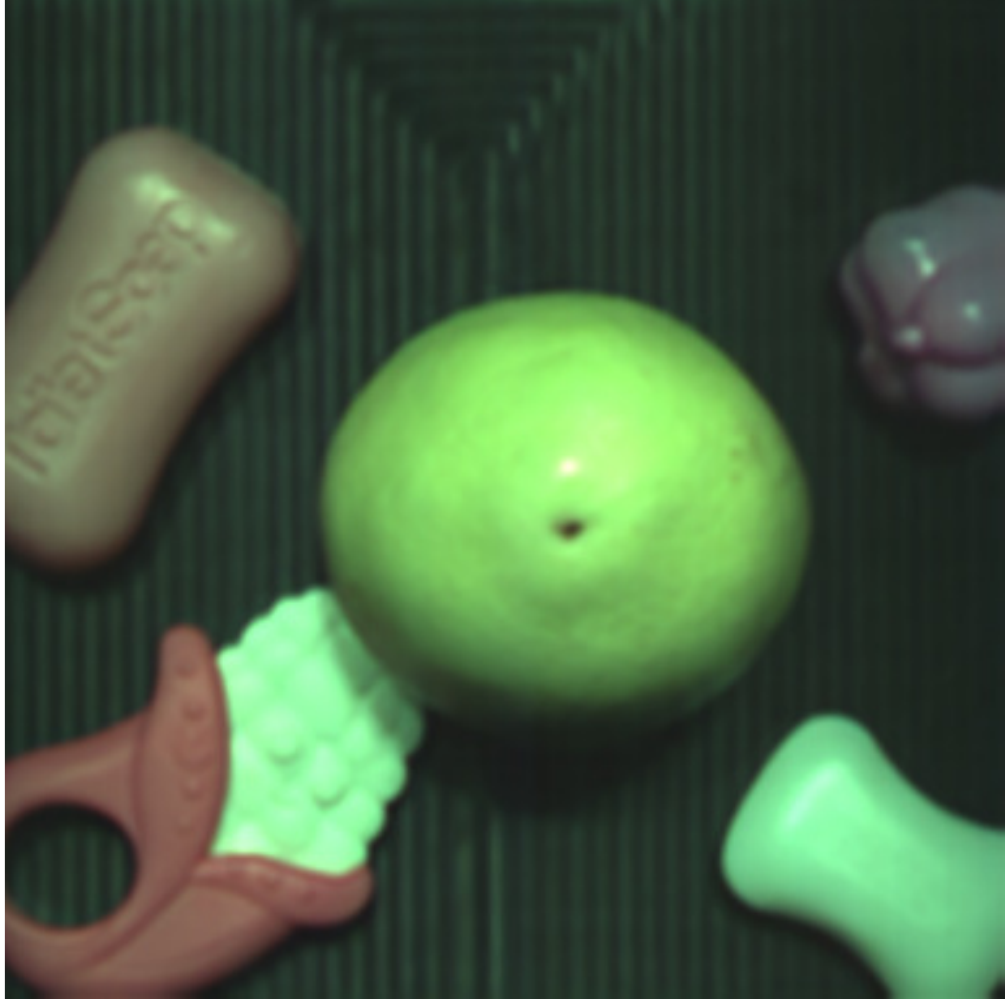}}
		\centerline{(h) ENCAM}\medskip
	\end{minipage}
	\hfill
	\begin{minipage}[b]{0.32\linewidth}
		\centering
		\centerline{\includegraphics[width=2.7cm,height=2.7cm]{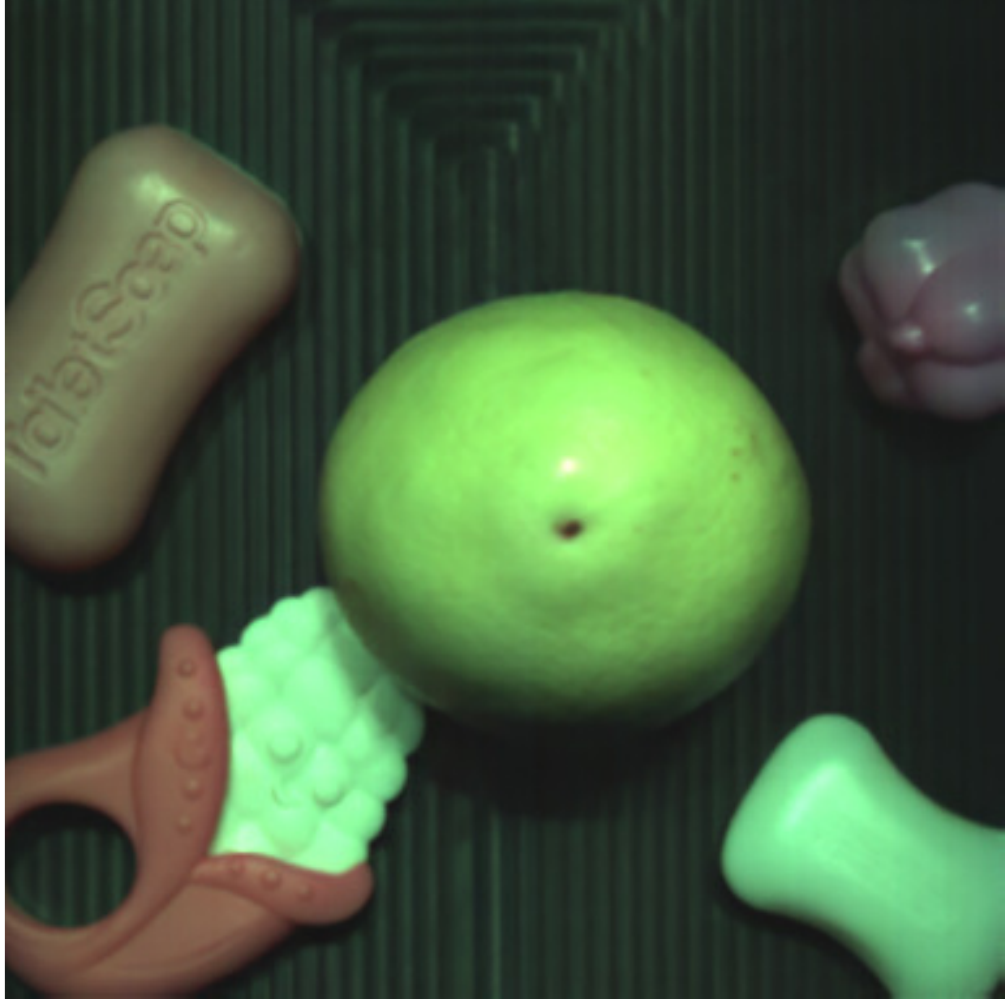}}
		\centerline{(i) HSIE (Ours)}\medskip
	\end{minipage}
	
	\vspace{-0.3cm}
	\caption
	{ Real low-light HSI enhancement results. Pseudo-color with bands (57, 27, 17).  (a) Low-light. (b) MR. (c) HE. (d) CLAHE. (e) MSR. (f) Retinex-Net. (g) 3D-ADNet. (h) ENCAM. (i) HSIE (Ours). Best viewed in color and zoomed in.}\medskip
	\label{fig:visual_comparision}  
	\vspace{-0.6cm}
\end{figure} 

The outdoor dataset contains 20 pairs of HSIs and is acquired using the SPECIM IQ\footnote{Detailed information about the SPECIM IQ can be found at \url{https://www.specim.fi/iq/}}, which is a compact scanning-based hyperspectral camera. To obtain an accurate image of the reflectance spectrum, we recalibrate the reflectance using a whiteboard whenever the lighting conditions change. The outdoor dataset is captured on campus and contains images of various objects, such as cars, trees, stones, and buildings. Each of the outdoor HSI has 204 bands with wavelengths ranging from 400nm to 1000nm and a spatial resolution of $512 \times 512$. For unification with indoor datasets, the first 6 bands and the last 6 bands are removed. Then, among the remaining 192 bands, we uniformly sample one band in every three bands, reserving about 64 bands. The final shape of the outdoor HSI is $512 \times512 \times64$. In the case of the outdoor dataset, we randomly chose 80 percent of the samples for training and the rest 20 percent for testing.

A training sample is created according to the following procedure. We first normalize all the HSIs in the range of [0,1]. Then, for each low-light HSI, a $64 \times64$ patch on each band without overlapping is cropped, and a $64 \times64 \times24$ cubic is cropped at the same position of its adjacent $24$ bands. Following this preprocessing procedure, 48 training samples can be cropped from an HSI with spatial resolution $390 \times 512$. The label is cropped from the same position as the corresponding long-exposure HSI using the same preprocessing method. In the end, 152,000 training samples are obtained on the indoor dataset and 20,480 training samples on the outdoor dataset.

\begin{figure}[htb]
	
	\begin{minipage}[b]{0.32\linewidth}
		\centering
		\centerline{\includegraphics[width=2.7cm,height=2.7cm]{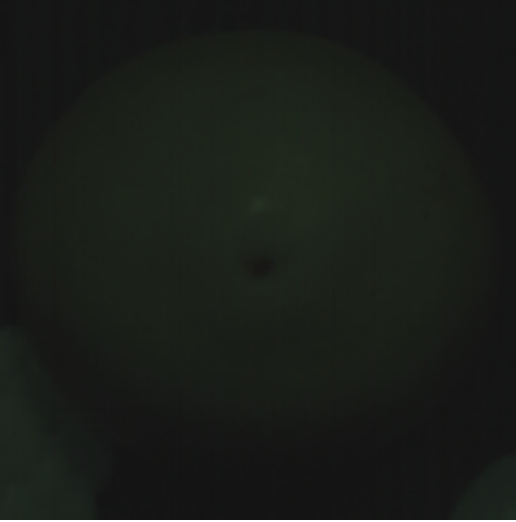}}
		\centerline{(a) Low-light}\medskip
	\end{minipage}
	\hfill
	\begin{minipage}[b]{0.32\linewidth}
		\centering
		\centerline{\includegraphics[width=2.7cm,height=2.7cm]{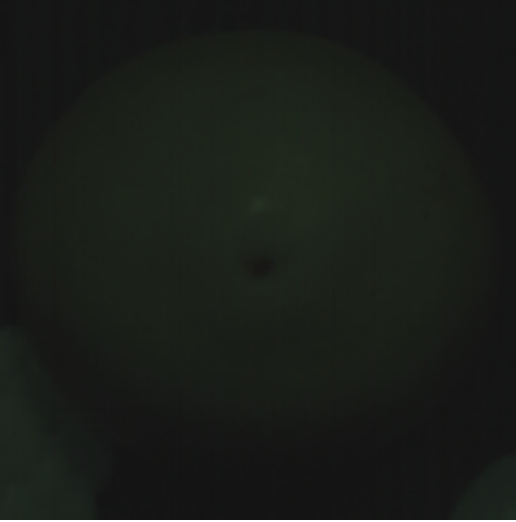}}
		\centerline{(b) MR}\medskip
	\end{minipage}
	\hfill
	\begin{minipage}[b]{0.32\linewidth}
		\centering
		\centerline{\includegraphics[width=2.7cm,height=2.7cm]{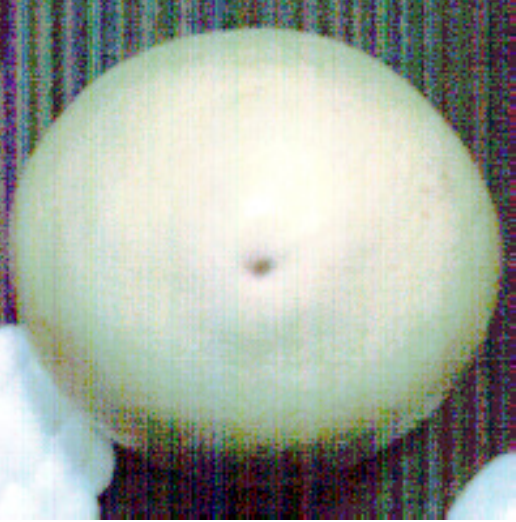}}
		\centerline{(c) HE}\medskip
	\end{minipage}
	\begin{minipage}[b]{0.32\linewidth}
		\centering
		\centerline{\includegraphics[width=2.7cm,height=2.7cm]{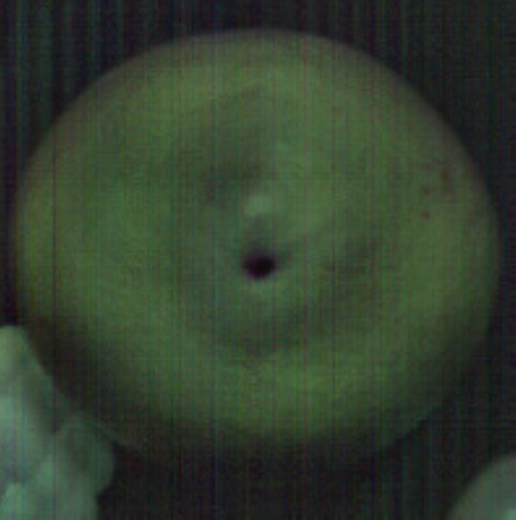}}
		\centerline{(d) CLAHE}\medskip
	\end{minipage}
	\hfill
	\begin{minipage}[b]{0.32\linewidth}
		\centering
		\centerline{\includegraphics[width=2.7cm,height=2.7cm]{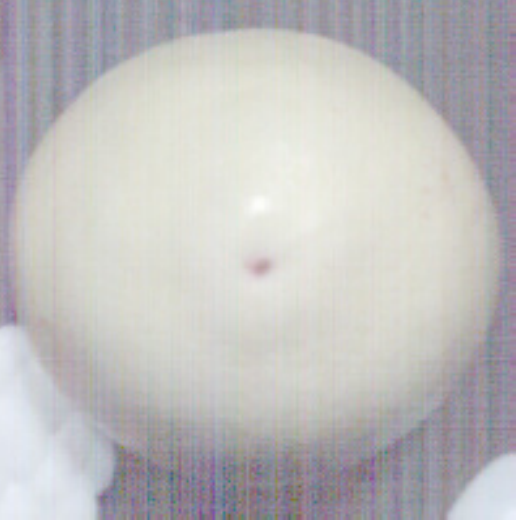}}
		\centerline{(e) MSR}\medskip
	\end{minipage}
	\hfill
	\begin{minipage}[b]{0.32\linewidth}
		\centering
		\centerline{\includegraphics[width=2.7cm,height=2.7cm]{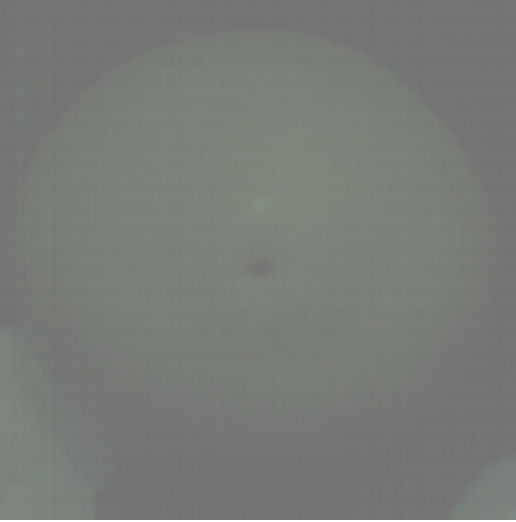}}
		\centerline{(f) Retinex-Net}\medskip
	\end{minipage}
	\begin{minipage}[b]{0.32\linewidth}
		\centering
		\centerline{\includegraphics[width=2.7cm,height=2.7cm]{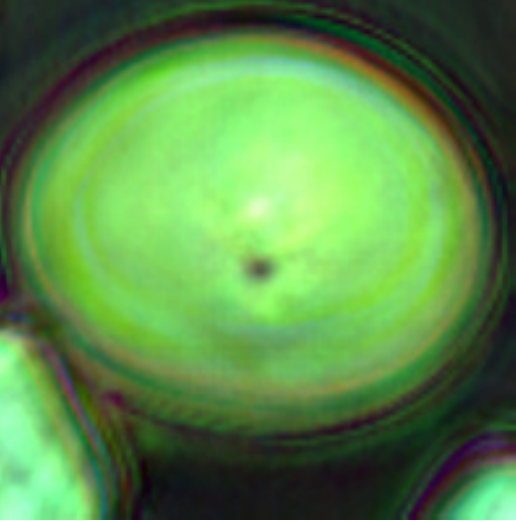}}
		\centerline{(g) 3D-ADNet}\medskip
	\end{minipage}
	\hfill
	\begin{minipage}[b]{0.32\linewidth}
		\centering
		\centerline{\includegraphics[width=2.7cm,height=2.7cm]{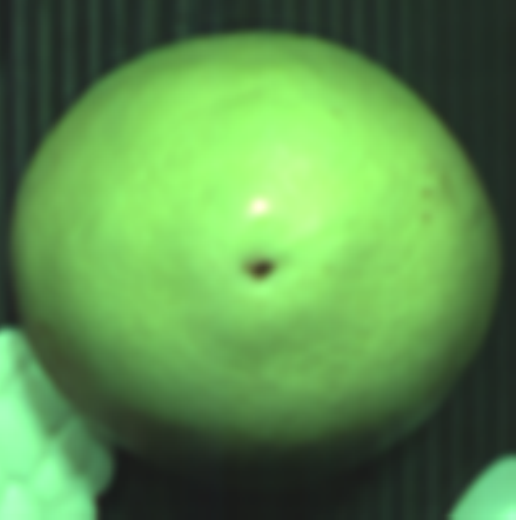}}
		\centerline{(h) ENCAM}\medskip
	\end{minipage}
	\hfill
	\begin{minipage}[b]{0.32\linewidth}
		\centering
		\centerline{\includegraphics[width=2.7cm,height=2.7cm]{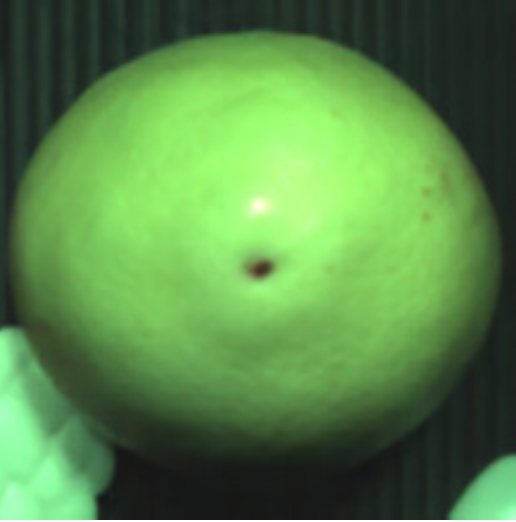}}
		\centerline{(i) HSIE (Ours)}\medskip
	\end{minipage}
	
	\vspace{-0.3cm}
	\caption
	{ Vision comparison of the highlighted region of interest. (a) Low-light. (b) MR. (c) HE. (d) CLAHE. (e) MSR. (f) Retinex-Net. (g) 3D-ADNet. (h) ENCAM. (i) HSIE (Ours). Best viewed in color and zoomed in.}\medskip
	\label{fig:roi_visual_comparision}  
	\vspace{-0.6cm}
\end{figure} 

\subsection{Implementation Details}

Our model is implemented using the prevalent Pytorch framework and is trained with a single NVIDIA GeForce RTX 3090 GPU. The training epoch is set as 600. Adam optimizer is adopted to optimize the proposed model, and the momentum parameters of the optimizer are set as $0.9$, $0.999$, and $10^{-8}$, respectively. We initialize the weights of the proposed model by the Kaiming initialization method\cite{kaiminghe_init}. The initial learning rate for training our model is set as $2 \times 10^{-4}$. We use a step learning rate scheduler that decreases in half for every 200 epochs. We take the $L_1$ loss function as the optimizing objective, which is proven to provide better convergence than the $L_2$\cite{lim2017enhanced}. In our case, $L_1$ loss also performs better than $L_2$ loss in terms of noise suppression. A detailed discussion of $L_1$ and $L_2$ loss is presented in Section \ref{sec:ablation_study}.

\subsection{Evalution Metrics}

We use three objective assessment measures to quantitatively assess the proposed HSIE's performance, including mean peak-signal-to-noise ratio (MPSNR)\cite{wang2004image}, mean structural similarity (MSSIM)\cite{wang2004image}, and spectral angle mapper (SAM)\cite{dennison2004comparison}. MPSNR and MSSIM are metrics for spatial features of HSI, SAM is used for evaluating spectral consistency. In general, better low-light HSI enhancement results are indicated by lower SAM and higher MPSNR, MSSIM values.

\subsection{Experiments on Indoor Dataset} \label{sec:experimental_on_indoor_dataset}

We compare the proposed HSIE with current mainstream HSI denoising methods such as ENCAM \cite{encam} 3D-ADNet\cite{3d_danet}. Also, the proposed method is compared with classic low-light natural image enhancement methods such as HE\cite{gonzalez2009digital}, CLAHE\cite{park2008contrast}, MSR\cite{jobson1997multiscale}, the McCann’s Retinex
(MR) method\cite{mccann1999lessons}, and  Retinex-Net\cite{retinexnet_wei2018deep}. In the MSR method, the scale parameters are 15, 80, and 360, respectively. In the MR method, we set the number of iterations to 3. By regarding each HSI band as an independent grayscale image, it is straightforward to utilize the aforementioned classic image enhancement methods to enhance the low-light HSI band. For a fair comparison, all deep-learning-based algorithms are trained on our collected LHSI dataset.

\begin{table}[t]
	\centering
	\footnotesize
	\renewcommand{\arraystretch}{1.2}
	\caption{Comparisons with mainstream HSIs methods on the indoor LHSI dataset.}
	
	\begin{tabular}{l|r|r|r}
		\hline
		Models & MRSNR$\uparrow$  & MSSIM$\uparrow$  & SAM$\downarrow$ \\
		\hline
		MR\cite{mccann1999lessons} & 13.498 & 0.6916 & 7.497 \\
		HE\cite{gonzalez2009digital} & 11.691 & 0.2840 & 24.064 \\
		CLAHE\cite{park2008contrast} & 16.924 & 0.6811 & 11.475 \\
		MSR\cite{jobson1997multiscale} & 7.282 & 0.4058 & 17.477 \\
		Retinex-Net\cite{retinexnet_wei2018deep} & 10.653 & 0.5273 & 13.543 \\
		3D-ADNet\cite{3d_danet} & 25.268 & 0.7325 & 8.0426 \\			
		ENCAM\cite{encam} & \underline{32.713} & \underline{0.9663} & \underline{2.3329} \\
		\hline
		\textbf{HSIE (Ours)} & \textbf{38.628} & \textbf{0.9794} & \textbf{1.3906} \\
		\hline
	\end{tabular}
	\smallskip
	
	\label{table:SOTA}
\end{table}

Table~\ref{table:SOTA} displays the quantitative evaluations of several comparison approaches. In Table~\ref{table:SOTA}, the top performance is emphasized in bold, while the second-best is underlined. In Table~\ref{table:SOTA}, the MPSNR value of 3D-ADNet method is 25.268, which is 13.360 dB smaller than the MPSNR value of the proposed approach. Further, as can be observed from Table~\ref{table:SOTA}, our approach HSIE performs favorably against the second-best ENCAM in all three metrics. Compared with ENCAM, the accuracy of our method in MPSNR and MSSIM are increased by 5.919 dB and 0.0131, respectively, while the SAM is decreased by 0.9423. The results from 3D-ADNet and ENCAM reveal that the current state-of-the-art HSI denoising approach is ineffective for low-light HSI enhancement. We conjecture that this is caused by ignoring the intrinsic properties of low-light HSI. In addition, we can observe that data-driven neural network methods outperform model-driven traditional algorithms by a significant margin.

Fig.~\ref{fig:visual_comparision} shows visual comparisons of the pseudo-color enhancement results of all the comparison methods. From Fig.~\ref{fig:visual_comparision}, we can see clearly that the model-driven image enhancement MR method is unable to enlighten the extreme dark areas of the HSI, which is consistent with the conclusion drawn from Table~\ref{table:SOTA}. The results of the HE and MSR show that the two methods can successfully enlighten the original image. However, they are unable to suppress the noise and keep spectral fidelity. CLAHE is also capable of enlightening dark areas. However, it introduces obvious textural distortion. The result of Retinex-Net demonstrates that it can enhance the low-light HSI to some extent, but the generated image looks unrealistic. From Fig.~\ref{fig:visual_comparision} (g), we can see clearly that the restoration result of the 3D-ADNet introduces heavy artifacts and looks blurry. The enhanced image of ENCAM loses some textural details and looks blurry too. This result can be highlighted by the observation in Fig.~\ref{fig:roi_visual_comparision} (h). Fig.~\ref{fig:roi_visual_comparision} demonstrates the magnified version of the highlighted red rectangle region of interest in Fig.~\ref{fig:visual_comparision} (a).  The result in Fig.~\ref{fig:roi_visual_comparision} (i) shows that the proposed approach gains the best result. It also demonstrates that the HSIE is capable of properly enlightening the dark regions of the low-light HSI while keeping textural details.

\begin{figure}[t]
	\centering
	\includegraphics[width=0.94\columnwidth]{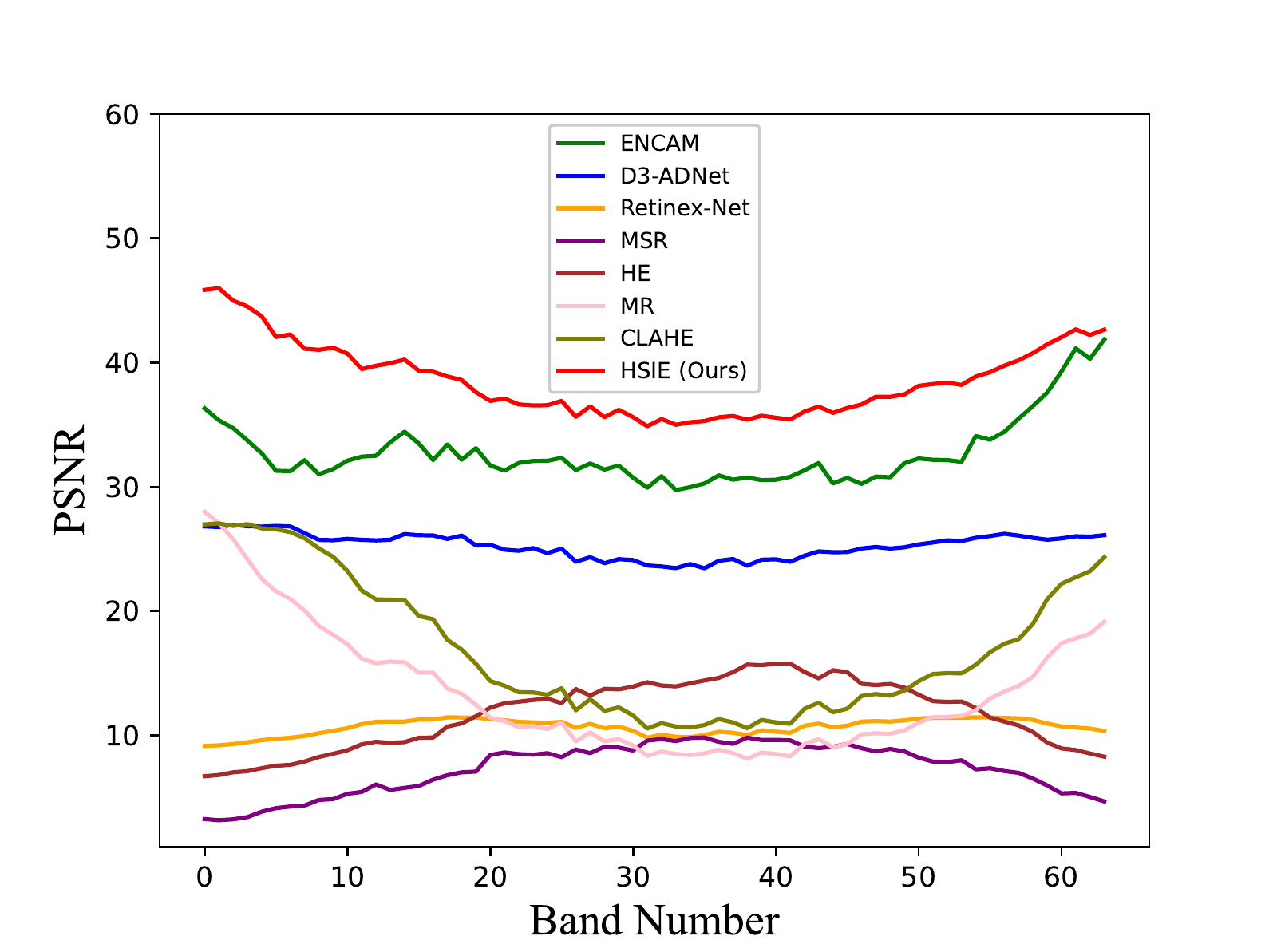}
	\caption{The PSNR curve is computed on a validation low-light HSI in the indoor LHSI dataset with different comparison methods. The proposed HSIE outperforms other comparison methods in all bands. }
	\label{fig:psnr_curve}
\end{figure}

To further demonstrate the superiority of our approach, Fig.~\ref{fig:psnr_curve} exhibits the PSNR curves evaluated on the validation low-light HSI with different comparison methods. One can observe that the proposed HSIE gains the top performance among all bands. The excellent performance attributes to all components of HSIE. For one thing, the multi-scale feature fusion mechanism in the shallow feature extraction module adequately extracted the joint spatial and spectral information. For another thing, the concatenated CABs show the benefits of enlightening the low-light band. Finally, the high-frequency refinement branch successfully adjusts textural details of low-light HSI.

The proposed model uses the current low-light band and its $k$ adjoining bands as inputs. We set the value of $k$ as 24 during all the training procedures on the indoor and outdoor LHSI datasets. A detailed discussion for the value of $k$ is provided in Section \ref{sec:ablation_study}. 

\begin{figure*}[h]
	
	\begin{minipage}[b]{0.32\linewidth}
		\centering
		\centerline{\includegraphics[width=1\linewidth]{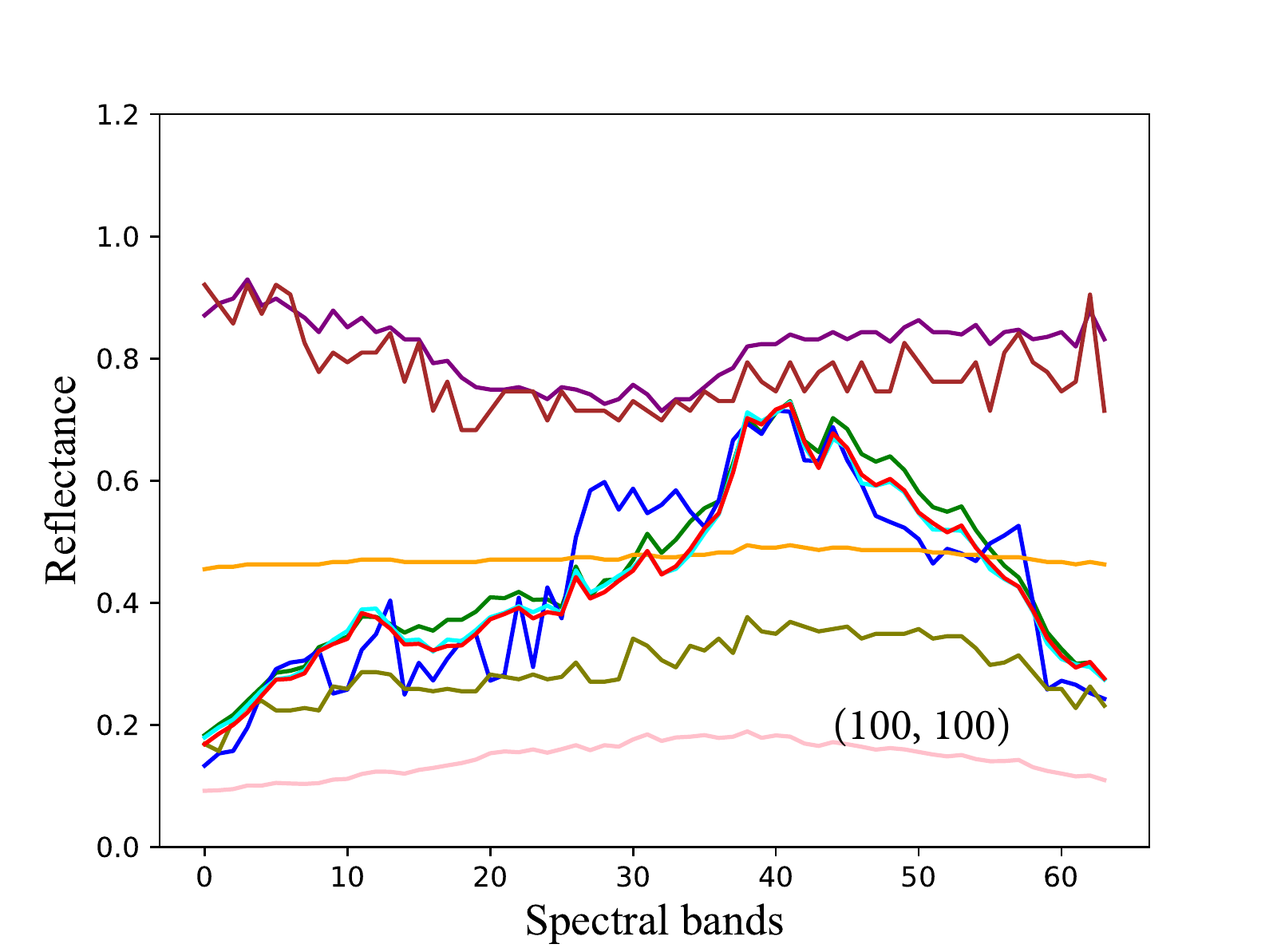}}
	\end{minipage}
	\hfill
	\begin{minipage}[b]{0.32\linewidth}
		\centering
		\centerline{\includegraphics[width=1\linewidth]{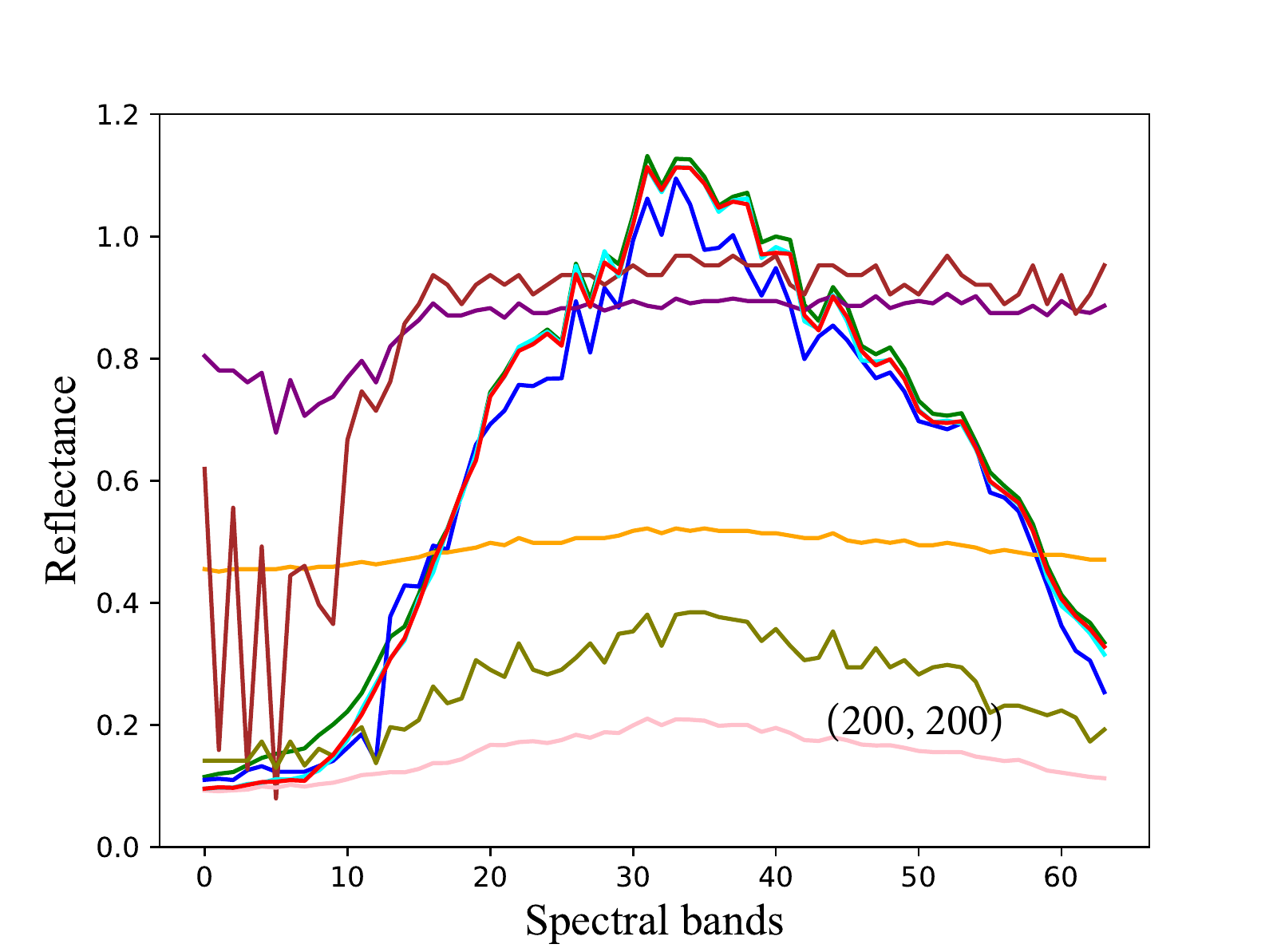}}
	\end{minipage}
	\hfill
	\begin{minipage}[b]{0.32\linewidth}
		\centering
		\centerline{\includegraphics[width=1\linewidth]{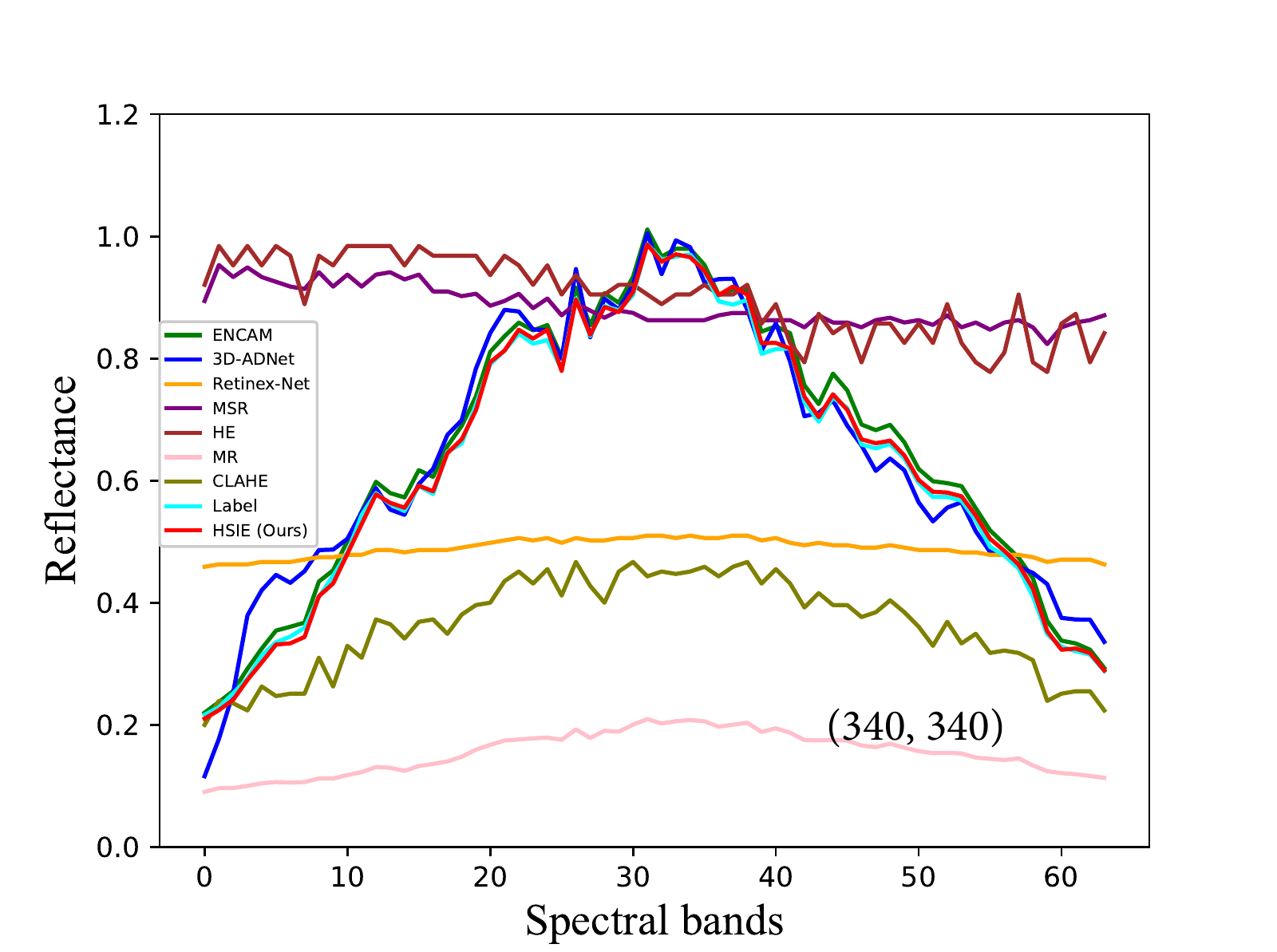}}
	\end{minipage}
	
	\vspace{-0.3cm}
	\caption
	{ Visual comparison of spectral distortion at pixel positions (100, 100), (200, 200), and (340, 340) on the validation HSI. The spectral curve generated by the proposed HSIE(in red) and by the label(in cyan) almost overlapped. Best viewed in color and zoomed in.}\medskip
	\label{fig:spectral_distortion}  
	\vspace{-0.2cm}
\end{figure*}

In addition, to show the variation of the spectral reflectance, the spectral distortion between the spectral curves in three positions obtained by the comparison models and ground truth is illustrated in Fig.~\ref{fig:spectral_distortion}. After whiteboard calibration, the cameras used can directly output both the DN (Digital Number) value and the reflectance. We select the reflectance for processing, so the curves illustrated in Fig.~\ref{fig:spectral_distortion} are computed based on spectral reflectance values. From Fig.~\ref{fig:spectral_distortion}, we can see that in most circumstances, the spectral curve generated by the proposed HSIE (red curve) overlaps the curve generated by the label (cyan curve) in all three positions. It implies that the proposed HSIE is the best to keep spectral fidelity.

\subsection{Experiments on Outdoor Dataset}

\begin{table*}[t]
	\centering
	\footnotesize
	\caption{Comparisons with mainstream HSIs methods on outdoor LHSI dataset.}
	\resizebox{0.95\linewidth}{!}{
		\begin{tabular}{l|rrr|rrr|rrr|rrr}
			\hline
			\multirow{2}{*}{Methods} & \multicolumn{3}{c|}{Car}    & \multicolumn{3}{c|}{Tree}     & \multicolumn{3}{c|}{Stone}     & \multicolumn{3}{c}{Building}     \\ \cline{2-13}
			& MPSNR$\uparrow$    & MSSIM$\uparrow$   & SAM$\downarrow$      & MPSNR$\uparrow$    & MSSIM$\uparrow$   & SAM$\downarrow$      & MPSNR$\uparrow$    & MSSIM$\uparrow$   & SAM$\downarrow$      & MPSNR$\uparrow$    & MSSIM$\uparrow$   & SAM$\downarrow$      \\
			\hline
			MR\cite{mccann1999lessons}       & 30.360       & 0.9402      & 4.630       & 24.304       & 0.8978      & 4.921       & 27.086       & 0.9043      & 3.8247       & 21.328       & 0.8079      & 6.003       \\
			LRMR\cite{LRMR}       & 29.052 & 0.9308 & 4.963  & 22.900 & 0.8788 & 4.404  & 25.354 & 0.8833 & 4.613  & 20.811 & 0.7903 & 6.988  \\
			HE\cite{gonzalez2009digital}         & 7.977  & 0.3677 & 17.708 & 8.755  & 0.4641 & 8.132  & 7.950  & 0.2905 & 17.292 & 8.862  & 0.4222 & 16.884 \\
			CLAHE\cite{park2008contrast}      & 32.609 & 0.9293 & 7.377  & 25.734 & 0.8295 & 6.680  & 28.973 & 0.8910 & 6.976  & 24.548 & 0.8587 & 9.111  \\
			MSR\cite{jobson1997multiscale}        & 4.428  & 0.2546 & 6.844  & 11.819 & 0.3581 & 11.555 & 8.032  & 0.2966 & 10.072 & 6.237  & 0.3629 & 10.924 \\
			Retinex-Net\cite{retinexnet_wei2018deep} & 7.384  & 0.4681 & 5.408  & 7.417  & 0.4328 & 4.605  & 7.453   & 0.4409 & 5.050   & 8.220   & 0.5032 & 7.409   \\
			ENCAM\cite{encam}      & 33.665  & \underline{0.9658}  & 4.757  & 24.373  & 0.8697  & 15.015  & 28.508  & \underline{0.9102}  & 11.898 & 23.636  & 0.8634  & 5.407 \\
			3D-ADNet\cite{3d_danet}       & \underline{39.801} & 0.9582 & \underline{2.093}  & \underline{27.491} & \underline{0.9013} & \underline{2.353}  & \underline{33.796} & 0.9054 & \underline{2.458}  & \textbf{26.621} & \underline{0.8849} & \textbf{2.408}  \\
			\hline
			HSIE (Ours)                & \textbf{46.840}  & \textbf{0.9882} & \textbf{0.816}  & \textbf{29.181} & \textbf{0.9412} & \textbf{2.027}  & \textbf{36.668} & \textbf{0.9576} & \textbf{1.359}   & \underline{26.565}  & \textbf{0.9239} & \underline{2.577}  \\
			\hline     
		\end{tabular}
		\smallskip
	} 
	\label{table:comparison_on_outdoor_dataset}
	\vspace{-0.1cm}
\end{table*}

\begin{figure}[htb]
	
	\begin{minipage}[b]{0.32\linewidth}
		\centering
		\centerline{\includegraphics[width=2.7cm,height=2.7cm]{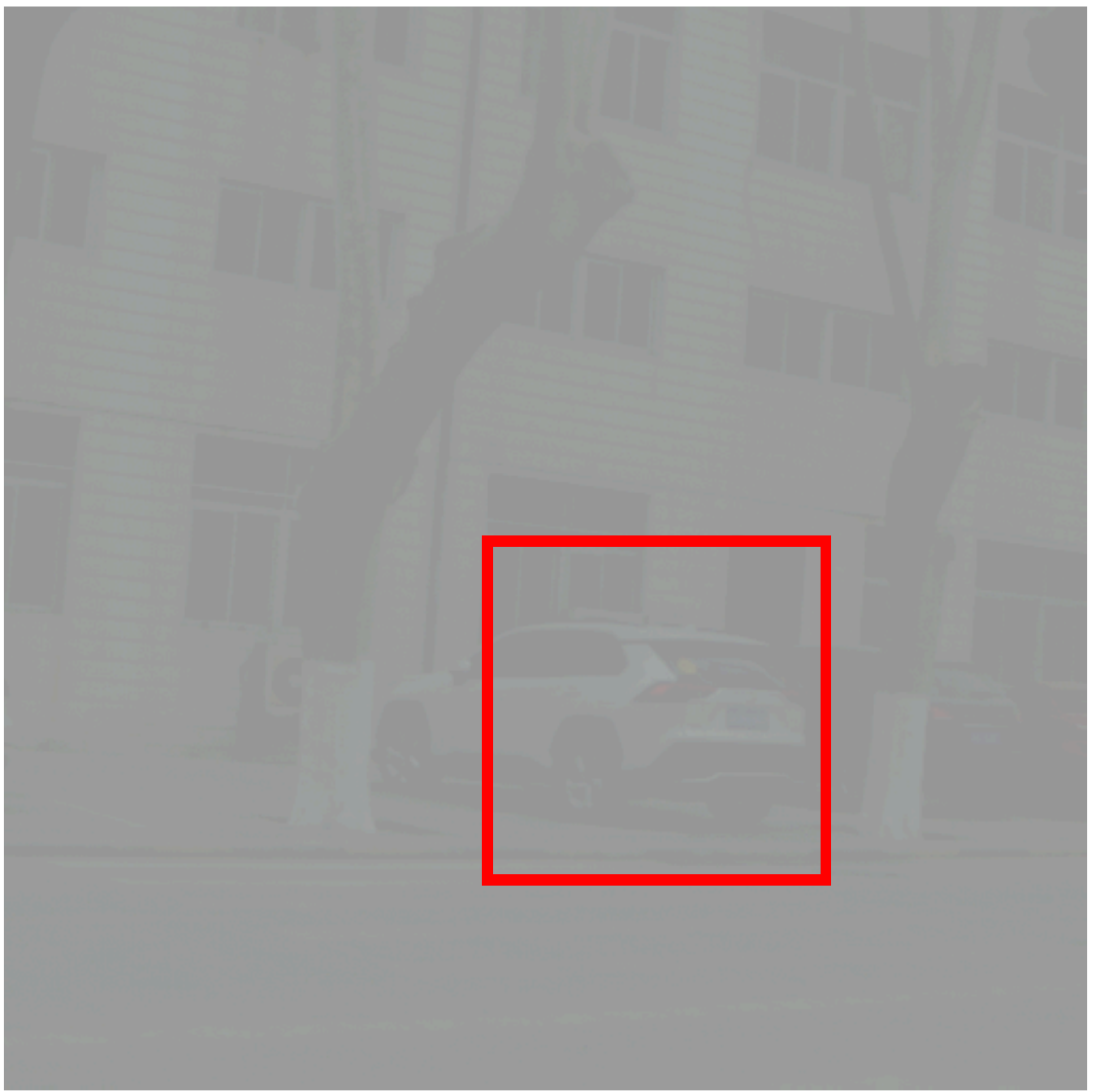}}
		\centerline{(a) Low-light$\ast$}\medskip
	\end{minipage}
	\hfill
	\begin{minipage}[b]{0.32\linewidth}
		\centering
		\centerline{\includegraphics[width=2.7cm,height=2.7cm]{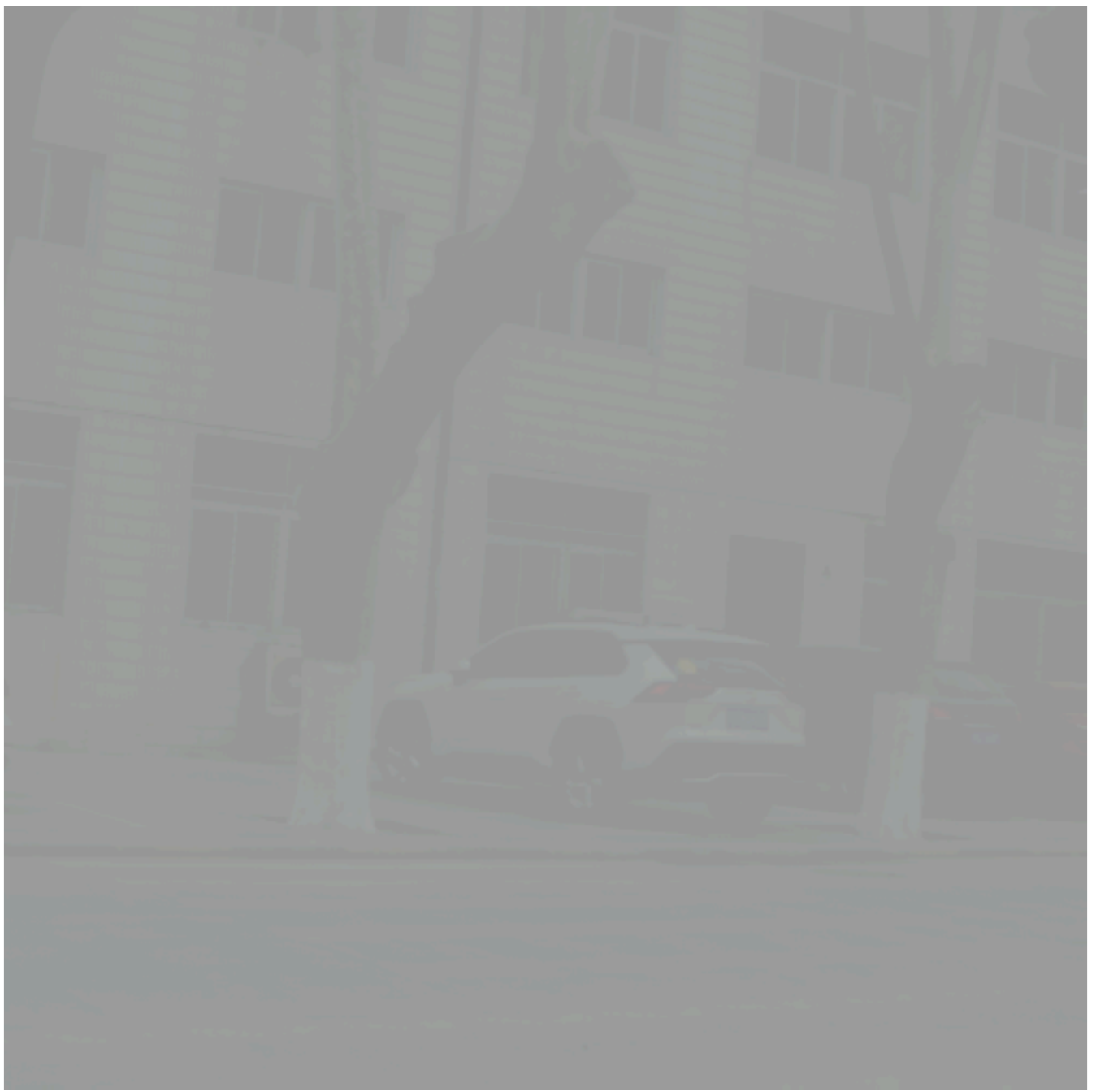}}
		\centerline{(b) LRMR$\ast$}\medskip
	\end{minipage}
	\hfill
	\begin{minipage}[b]{0.32\linewidth}
		\centering
		\centerline{\includegraphics[width=2.7cm,height=2.7cm]{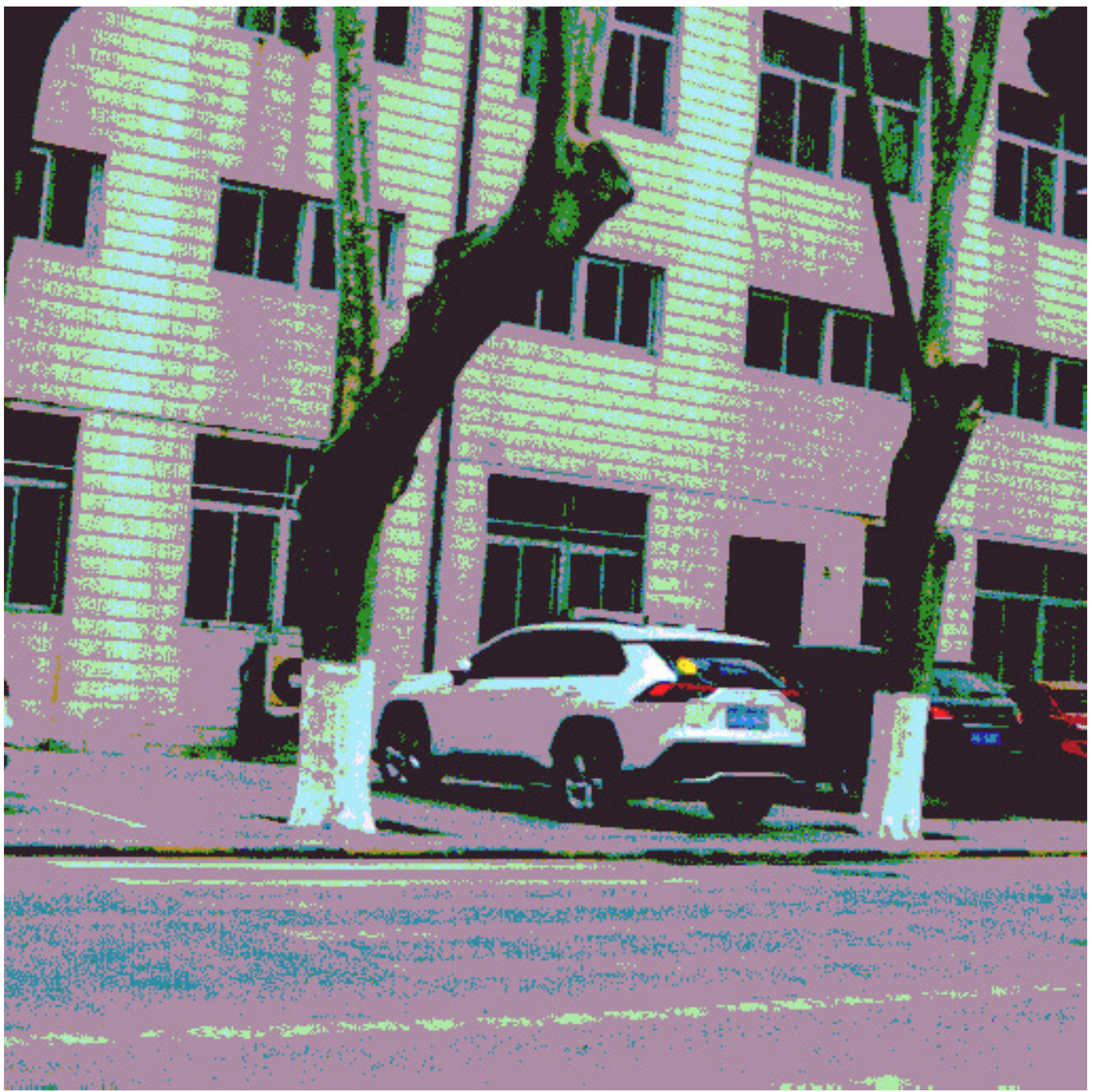}}
		\centerline{(c) HE}\medskip
	\end{minipage}
	\begin{minipage}[b]{0.32\linewidth}
		\centering
		\centerline{\includegraphics[width=2.7cm,height=2.7cm]{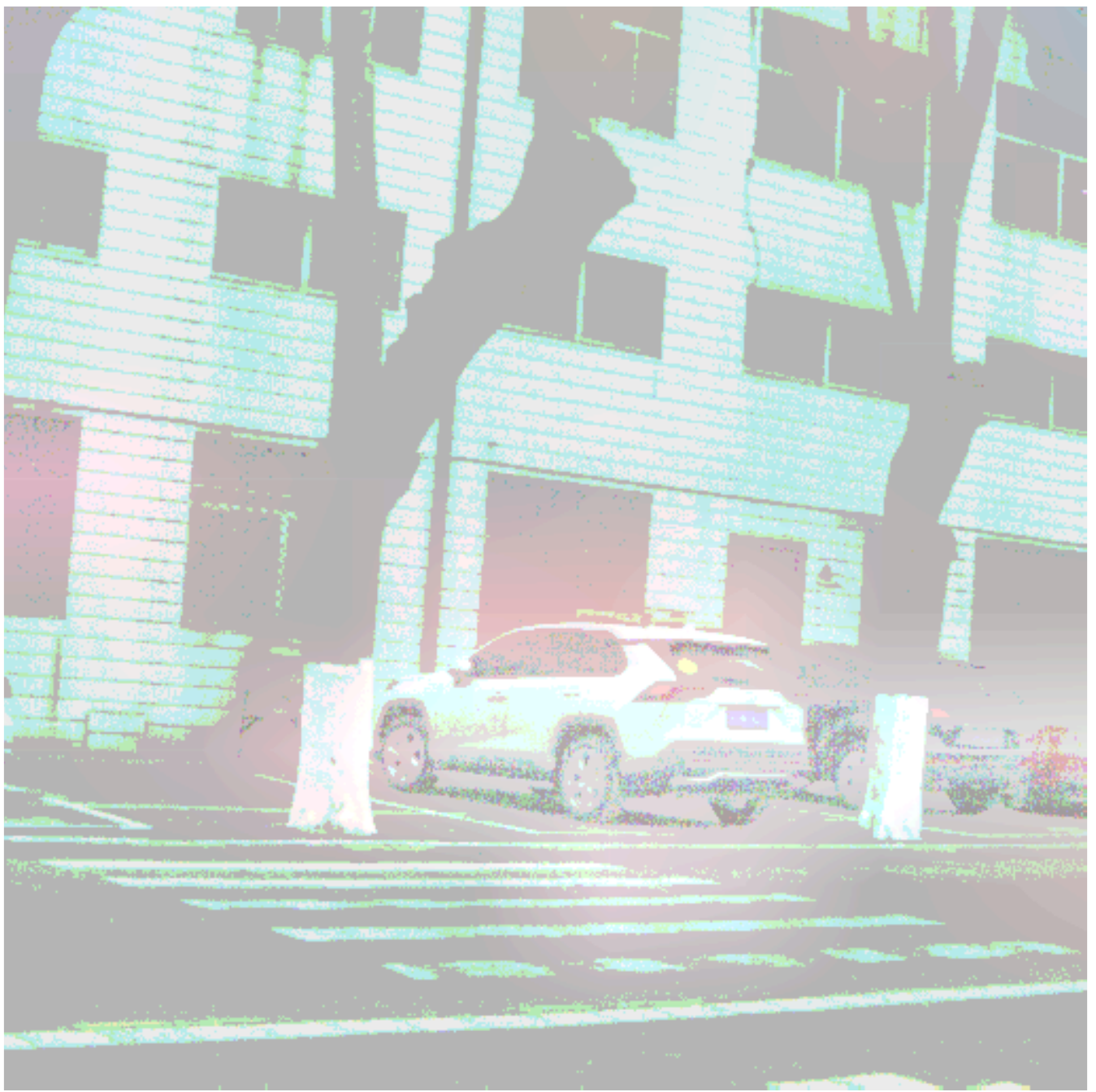}}
		\centerline{(d) CLAHE$\ast$}\medskip
	\end{minipage}
	\hfill
	\begin{minipage}[b]{0.32\linewidth}
		\centering
		\centerline{\includegraphics[width=2.7cm,height=2.7cm]{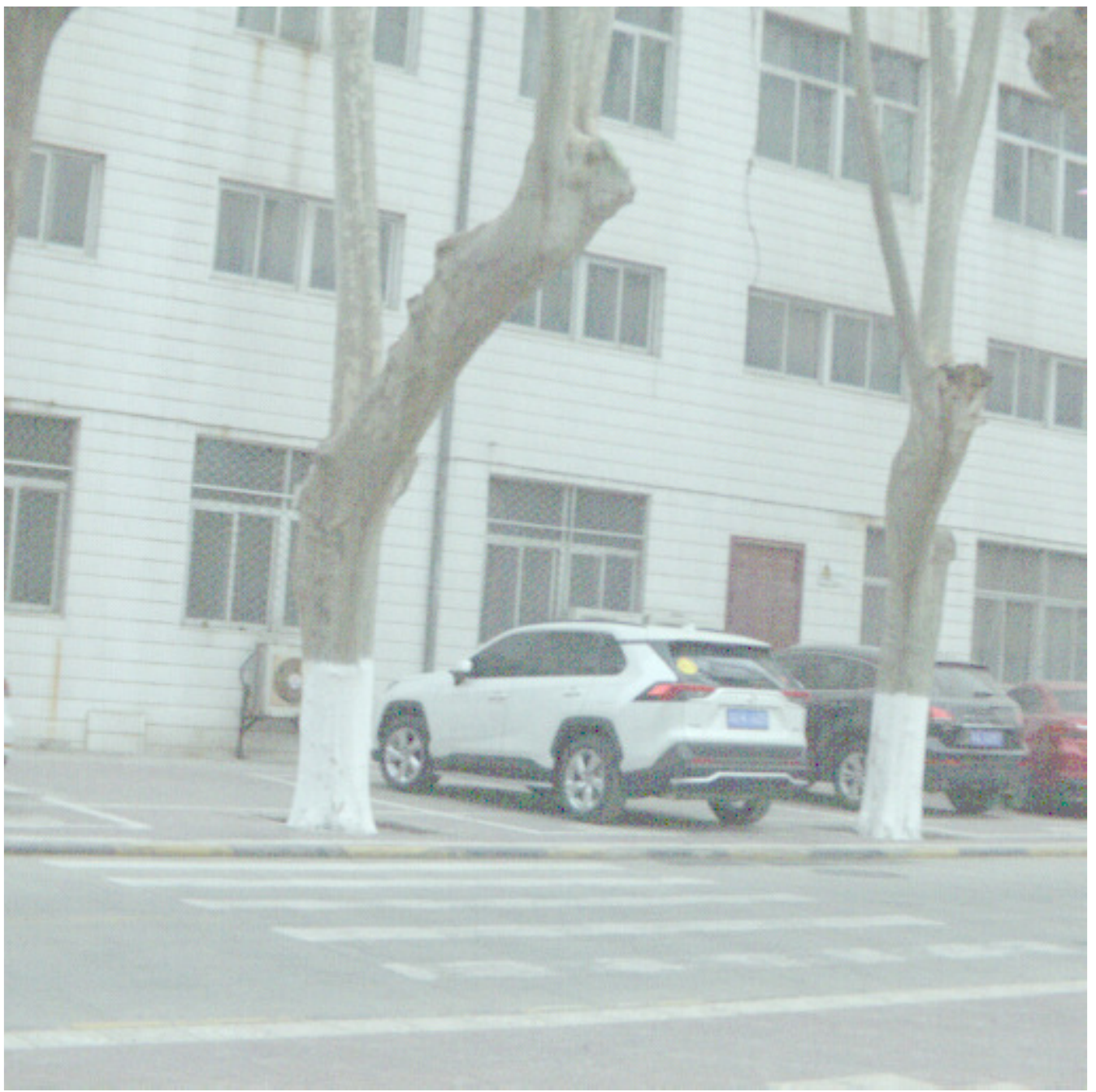}}
		\centerline{(e) MSR}\medskip
	\end{minipage}
	\hfill
	\begin{minipage}[b]{0.32\linewidth}
		\centering
		\centerline{\includegraphics[width=2.7cm,height=2.7cm]{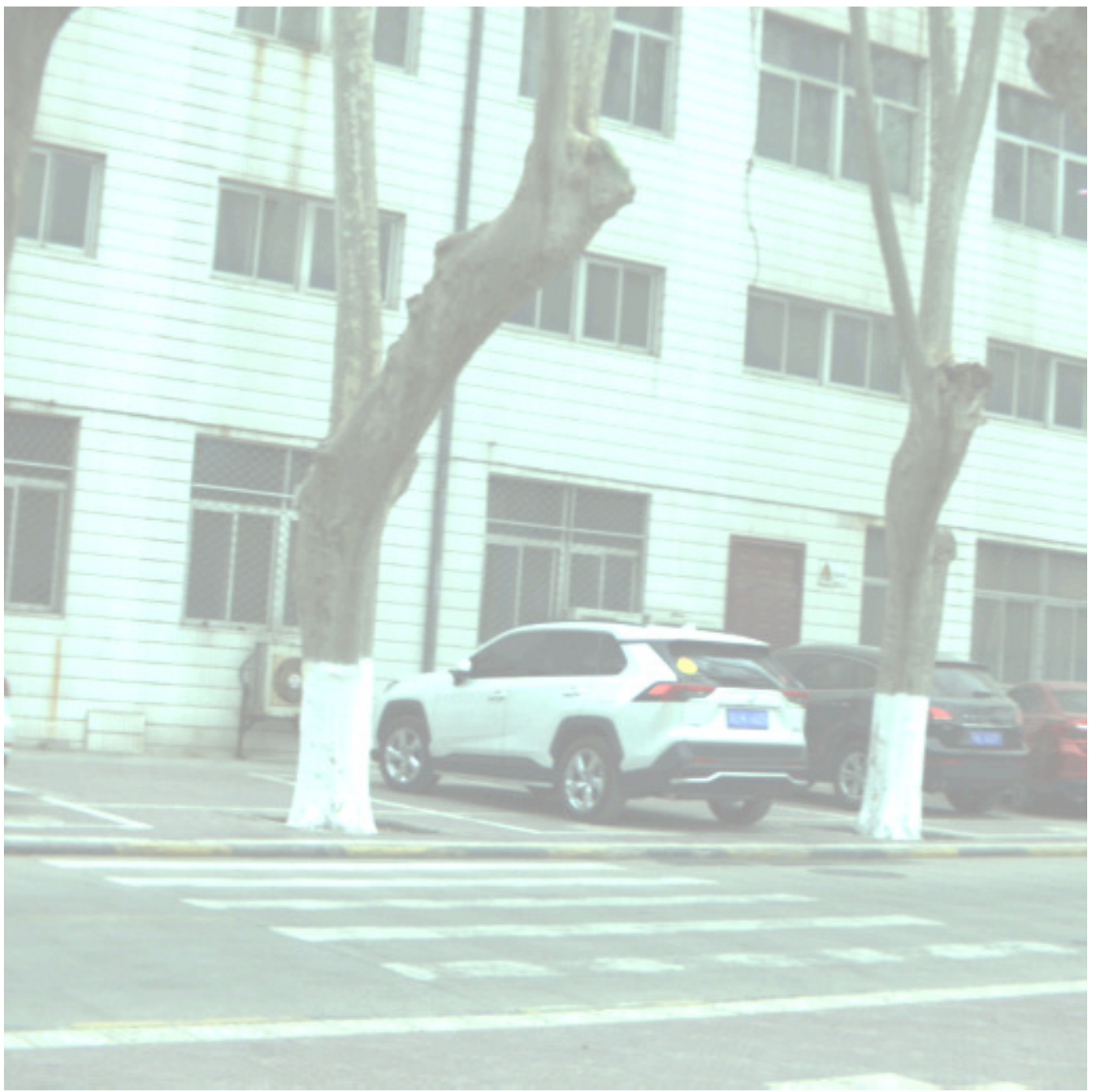}}
		\centerline{(f) Label$\ast$}\medskip
	\end{minipage}
	
	\begin{minipage}[b]{0.32\linewidth}
		\centering
		\centerline{\includegraphics[width=2.7cm,height=2.7cm]{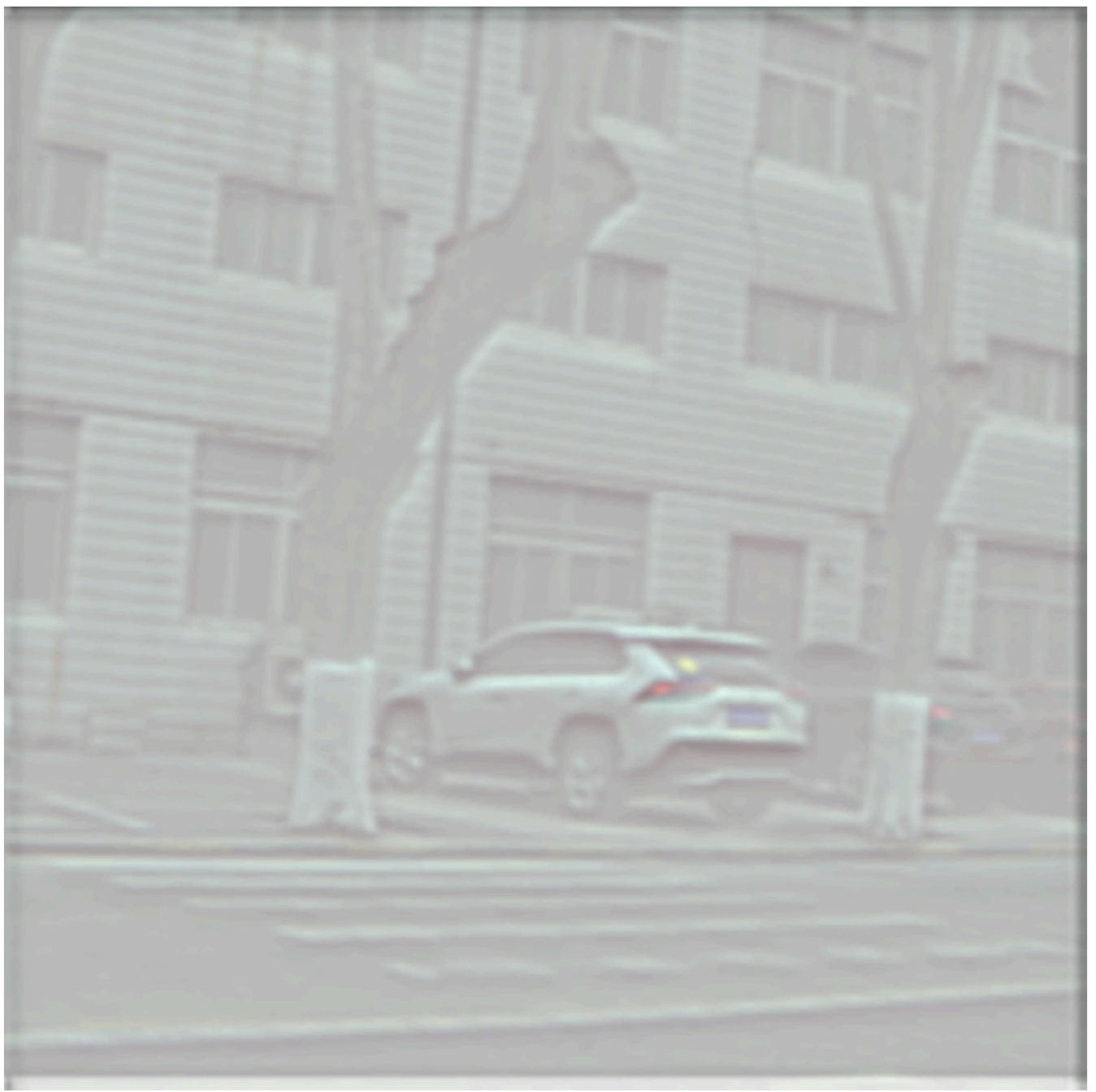}}
		\centerline{(g) ENCAM$\ast$}\medskip
	\end{minipage}
	\hfill
	\begin{minipage}[b]{0.32\linewidth}
		\centering
		\centerline{\includegraphics[width=2.7cm,height=2.7cm]{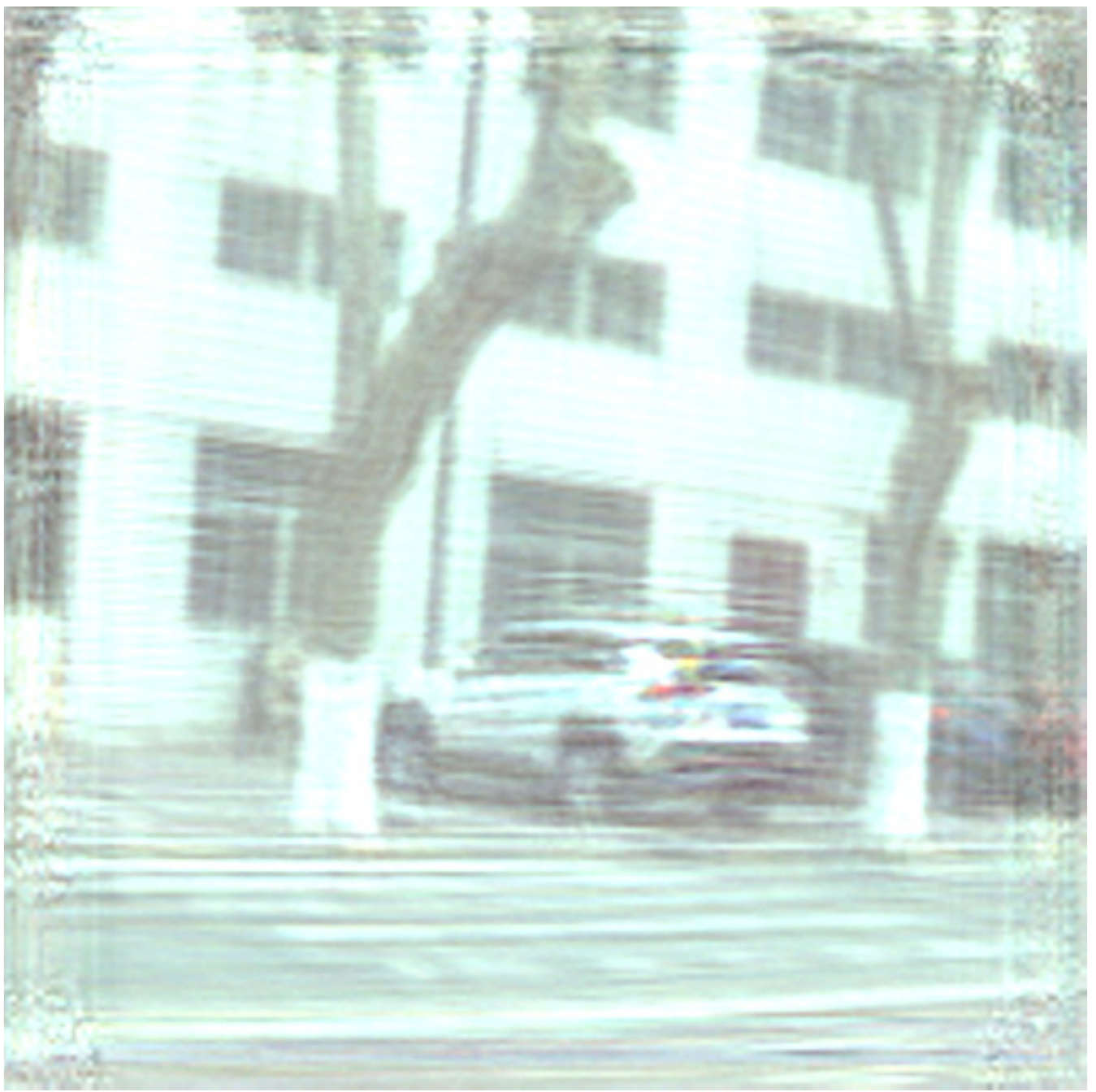}}
		\centerline{(h) 3D-ADNet$\ast$}\medskip
	\end{minipage}
	\hfill
	\begin{minipage}[b]{0.32\linewidth}
		\centering
		\centerline{\includegraphics[width=2.7cm,height=2.7cm]{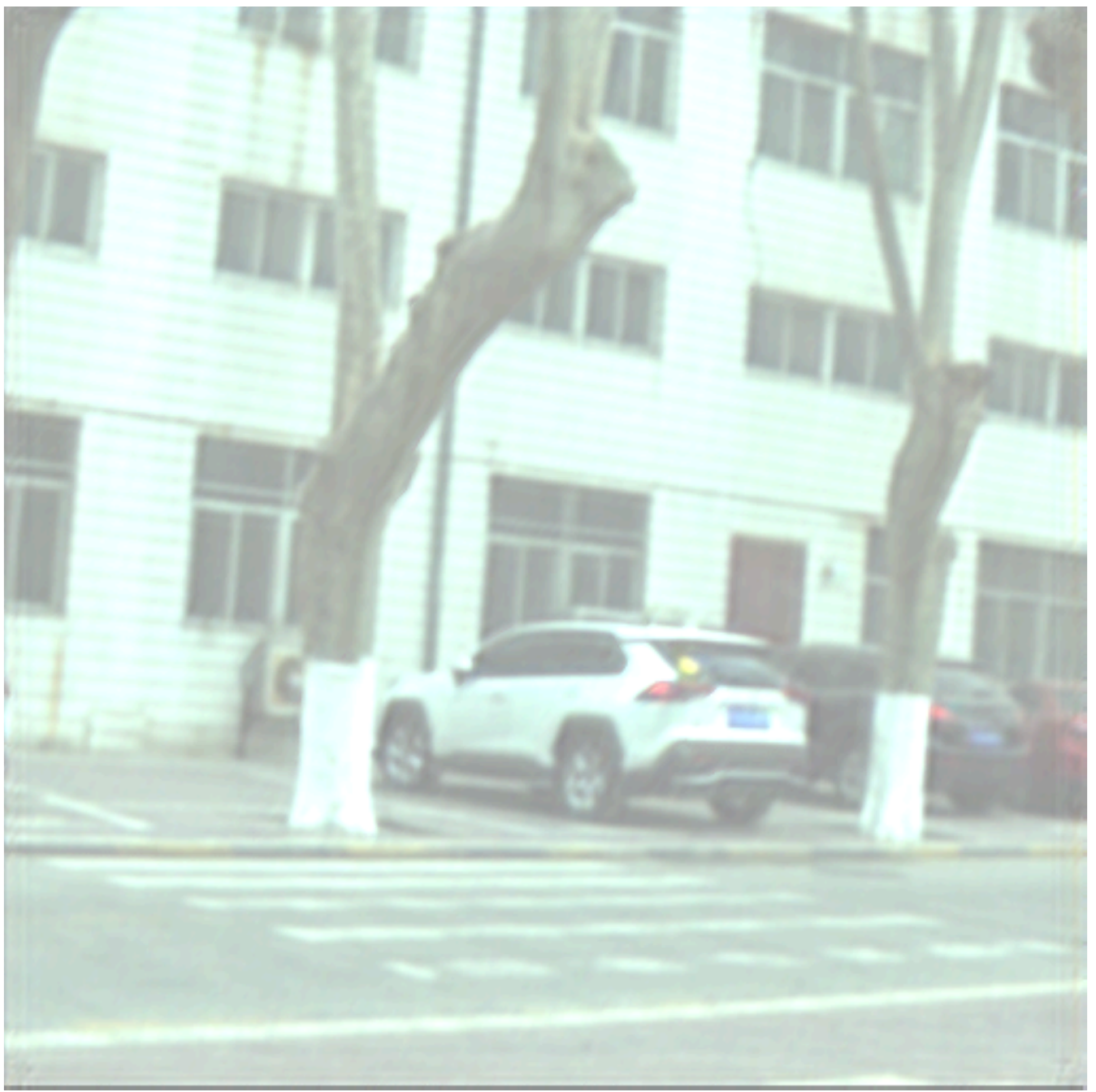}}
		\centerline{(i) HSIE (Ours)$\ast$}\medskip
	\end{minipage}
	
	\vspace{-0.3cm}
	\caption
	{ Vision comparison of the pseudo-color enhanced results on the low-light HSI. Band 57, 27, and 17 are selected to simulate red, green, and blue, respectively. (a) Low-light. (b) LRMR. (c) HE. (d) CLAHE. (e) MSR. (f) Label. (g) ENCAM. (h) 3D-ADNet. (i) HSIE (Ours). $\ast$ denotes that the image is linearly stretched for better viewing. Best viewed in color and zoomed in.}\medskip
	\label{fig:outdoor_visual_comparision}  
	\vspace{-0.6cm}
\end{figure} 

\begin{figure}[htb]
	
	\begin{minipage}[b]{0.32\linewidth}
		\centering
		\centerline{\includegraphics[width=2.7cm,height=2.7cm]{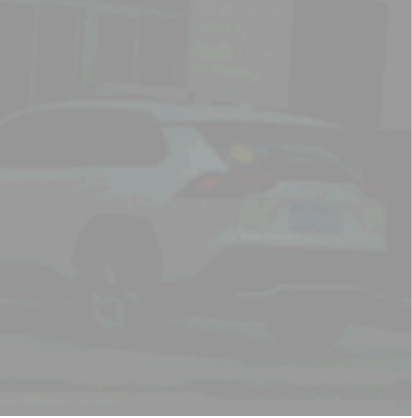}}
		\centerline{(a) Low-light$\ast$}\medskip
	\end{minipage}
	\hfill
	\begin{minipage}[b]{0.32\linewidth}
		\centering
		\centerline{\includegraphics[width=2.7cm,height=2.7cm]{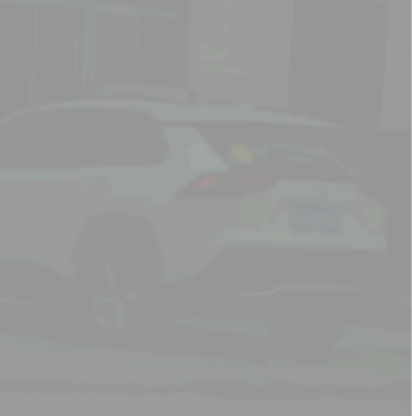}}
		\centerline{(b) LRMR$\ast$}\medskip
	\end{minipage}
	\hfill
	\begin{minipage}[b]{0.32\linewidth}
		\centering
		\centerline{\includegraphics[width=2.7cm,height=2.7cm]{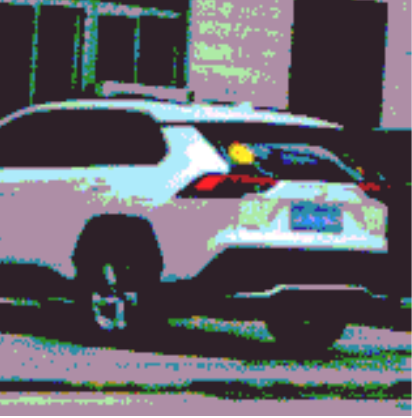}}
		\centerline{(c) HE}\medskip
	\end{minipage}
	\begin{minipage}[b]{0.32\linewidth}
		\centering
		\centerline{\includegraphics[width=2.7cm,height=2.7cm]{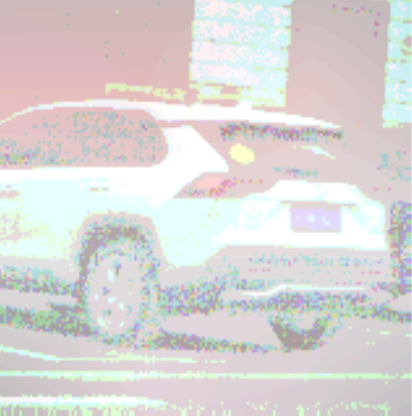}}
		\centerline{(d) CLAHE$\ast$}\medskip
	\end{minipage}
	\hfill
	\begin{minipage}[b]{0.32\linewidth}
		\centering
		\centerline{\includegraphics[width=2.7cm,height=2.7cm]{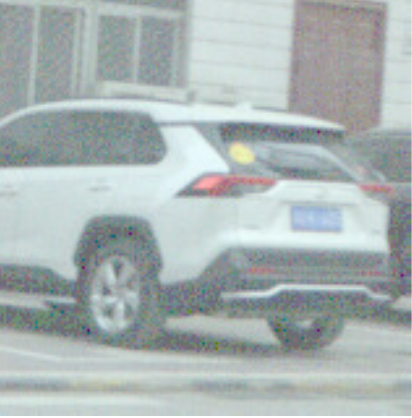}}
		\centerline{(e) MSR}\medskip
	\end{minipage}
	\hfill
	\begin{minipage}[b]{0.32\linewidth}
		\centering
		\centerline{\includegraphics[width=2.7cm,height=2.7cm]{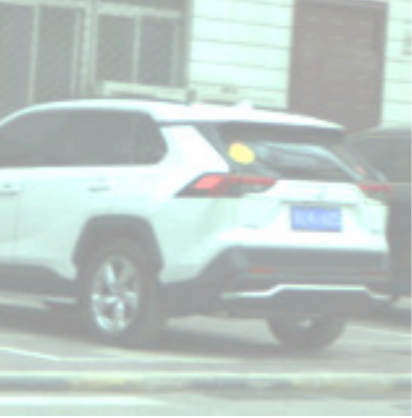}}
		\centerline{(f) Label$\ast$}\medskip
	\end{minipage}
	
	\begin{minipage}[b]{0.32\linewidth}
		\centering
		\centerline{\includegraphics[width=2.7cm,height=2.7cm]{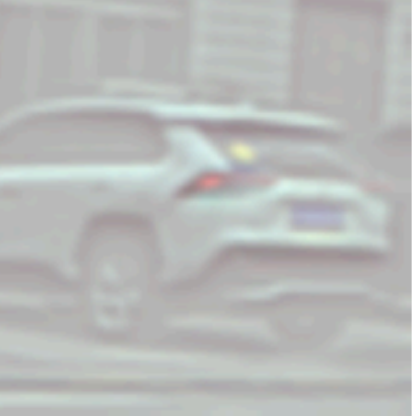}}
		\centerline{(g) ENCAM$\ast$}\medskip
	\end{minipage}
	\hfill
	\begin{minipage}[b]{0.32\linewidth}
		\centering
		\centerline{\includegraphics[width=2.7cm,height=2.7cm]{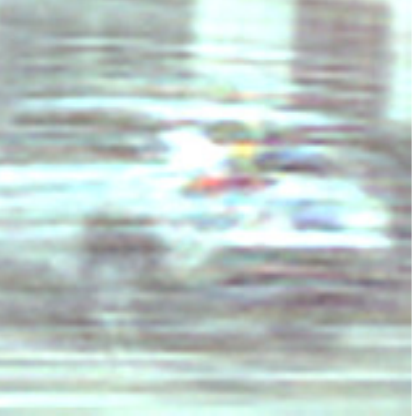}}
		\centerline{(h) 3D-ADNet$\ast$}\medskip
	\end{minipage}
	\hfill
	\begin{minipage}[b]{0.32\linewidth}
		\centering
		\centerline{\includegraphics[width=2.7cm,height=2.7cm]{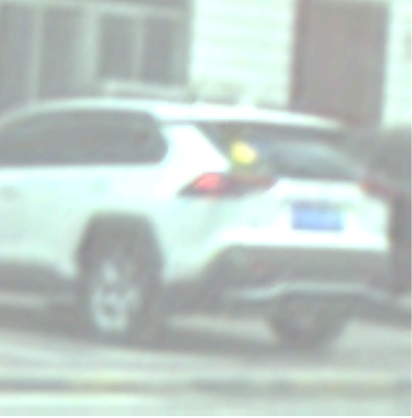}}
		\centerline{(i) HSIE (Ours)$\ast$}\medskip
	\end{minipage}
	
	\vspace{-0.3cm}
	\caption
	{ Vision comparison of the highlighted region of interest. (a) Low-light. (b) LRMR. (c) HE. (d) CLAHE. (e) MSR. (f) Label. (g) ENCAM. (h) 3D-ADNet. (i) HSIE (Ours). $\ast$ denotes that the image is linearly stretched for better viewing. Best viewed in color and zoomed in.}\medskip
	\label{fig:roi_visual_comparision_on_outdoor_dataset}  
	\vspace{-0.6cm}
\end{figure} 

The collected outdoor LHSI dataset is employed to support the training and testing of HSIE to further verify its efficacy. The enhancement results of the specific outdoor low-light HSI using different algorithms are demonstrated in Fig.~\ref{fig:outdoor_visual_comparision} for visual comparison. Fig.~\ref{fig:roi_visual_comparision_on_outdoor_dataset} demonstrates the magnified version of the highlighted red rectangle region of interest in Fig.~\ref{fig:outdoor_visual_comparision} (a). For better viewing, the darker images in Fig.~\ref{fig:outdoor_visual_comparision} and Fig.~\ref{fig:roi_visual_comparision_on_outdoor_dataset} are linearly stretched using the Matlab \emph{imadjust}\footnote{\url{https://ww2.mathworks.cn/help/images/ref/imadjust.html?lang=en}} function with consistent parameters. Unless specifically mentioned, images marked with an asterisk in this article denote that they are linearly stretched. From Fig.~\ref{fig:outdoor_visual_comparision}, we can draw a similar conclusion that the traditional HSI denoising algorithm LRMR is unable to enhance low-light HSIs, while deep-learning-based HSI denoising methods ENCAM and 3D-ADNet can successfully enlighten the darkened HSIs. However, the enhanced result of ENCAM looks darker. This result demonstrates that ENCAM can not improve the illumination sufficiently. The enhanced result of 3D-ADNet introduces some strip artifacts on the whole image (see Fig.~\ref{fig:roi_visual_comparision_on_outdoor_dataset} (h)), indicating that 3D-ADNet can not restore the textural details. These results once again proved that HSI denoising methods are not the perfect solutions for the low-light HSI enhancement problems. The most straightforward reason is that HSI denoising methods are not designed for low-light HSI enhancement. We can generate another conclusion from  Fig.~\ref{fig:outdoor_visual_comparision} and Fig.~\ref{fig:roi_visual_comparision_on_outdoor_dataset}, where model-driven natural image enhancement methods (such as HE, CLAHE and MSR) are competent to boost the illumination of low-light HSI. However, these methods only focus on improving the illumination, resulting in high brightness while ignoring the preservation of spectral fidelity and noise suppression. Finally, when compared to other competing approaches, the proposed HSIE produced the best visual results.

We also list the quantitative results evaluated with the aforementioned metrics on four scenes in the outdoor LHSI dataset in Table~\ref{table:comparison_on_outdoor_dataset},  where for each metric the top performance is highlighted in bold and the sub-optimal is underlined. The four scenes are with different light conditions and objects of various materials, which comprise stone, tree, building, and car, as demonstrated in Fig~\ref{fig:show_dataset_sample} (c), (d), (e), and (f), respectively. As shown in Table~\ref{table:comparison_on_outdoor_dataset}, the top performance in all three metrics is achieved by the proposed HSIE in almost all four scenes, demonstrating that the proposed HSIE can generalize well in different light conditions. 

\subsection{Ablation Study} \label{sec:ablation_study}
In this section, we first examine each of the HSIE module's effectiveness, then we investigate the impact of some key model design hyper-parameters, such as the adjoining spectral band number, and the CAB number in the enlightening module, \emph{etc.} Finally, the results of the analysis of model complexity and the impact of different loss functions are demonstrated. Unless otherwise specified, the ablation experiments in this section are performed on the indoor LHSI datasets.

\begin{table}[htb]
	\centering
	\small
	\setlength{\tabcolsep}{0.95mm}
	
	\caption{Ablation study of different component of HSIE. }
	\begin{tabular}{|c|c|c|c|}
		\hline
		Evaluation metric & MPSNR$\uparrow$           & MSSIM$\uparrow$          & SAM$\downarrow$          \\ \hline
		RDN               & 37.278                   & 0.9727                   & 2.061                    \\ \hline
		RDN+SFE          & 37.494                   & 0.9737                   & 1.958                    \\ \hline
		RDN+SFE+CAB      & 37.699                   & 0.9746                   & 1.749                   \\ \hline
		RDN+SFE+CAB+HIGH & \textbf{38.628} & \textbf{0.9794} & \textbf{1.390} \\ \hline
	\end{tabular}
	\vspace{-0.15cm}
	\label{table:module_ablation}  
\end{table}

The results of the ablation investigation into the effects of each module are reported in Table~\ref{table:module_ablation}. A simplified residual dense network (RDN)\cite{zhang2020residual} is adopted as our initial backbone, which is considered to be a very efficient module for image restoration.  We observe that "RDN+SFE" outperforms "RDN", which shows the benefit of using multi-scale Shallow Feature Extraction (SFE). Another observation is that "RDN+SFE+CAB" performs more favorably than "RDN+SFE", which indicates the advantage of employing CAB. Finally, our full approach "RDN+SFE+CAB+HIGH" achieves the best results on the LHSI dataset, where "HIGH" stands for the high-frequency refinement branch. The results in Table~\ref{table:module_ablation} show that each module can enhance restoration performance significantly since each module has a positive effect on information and gradient flow. Furthermore, the CAB can make use of cross-channel interaction to adaptively inhibit less useful features and stress significant features.

For the rest of this section, we concentrate on investigations into some key model design hyper-parameters.
The adjoining spectral band number $k$ plays a vital role in the whole enhancement process. Hence, we investigate $k$ to evaluate its influence. In table~\ref{table:study_of_adj_bands}, we demonstrate the investigation result of adjacent spectral band number $k$. The best performance is highlighted in bold. When the value of $k$ is set as $24$, our model gains the best performance evaluated with MPSNR. In the meanwhile, the performance of the proposed HSIE decreases when the value of $k$ is larger or smaller than $24$. We conjecture that the intrinsic association between distinct bands of an HSI is primarily responsible. The adjoining 24 bands of a single band are already sufficient to encode all the correlations along with spectral information. Less adjoining bands cause loss of information, while more adjoining bands introduce noisy information.

\begin{table}[htb]
	\centering
	\footnotesize
	\renewcommand{\arraystretch}{1.2}
	\caption{Study of adjacent spectral band number $k$.}
	
	\begin{tabular}{l|c|c|c}
		\hline
		Adjacent Band Number ($k$) & MPSNR$\uparrow$  & MSSIM$\uparrow$  & SAM$\downarrow$ \\
		\hline
		$k = 8$ & 37.466 & 0.9718 & 1.660 \\
		$k = 12$ & 37.885 & 0.9754 & 1.662 \\
		$k = 18$ & 37.842 & 0.9758 & 1.666 \\
		\bm{$k = 24$} \textbf{(Ours)}& \textbf{38.628} & \textbf{0.9794} & \textbf{1.390} \\
		$k = 36$ & 37.175 & 0.9722 & 1.868 \\	
		\hline
	\end{tabular}
	\smallskip
	\label{table:study_of_adj_bands}
\end{table}

\begin{figure*}[ht]
	\centering
	\begin{center}
		\subfloat{\includegraphics[width=0.32\textwidth]{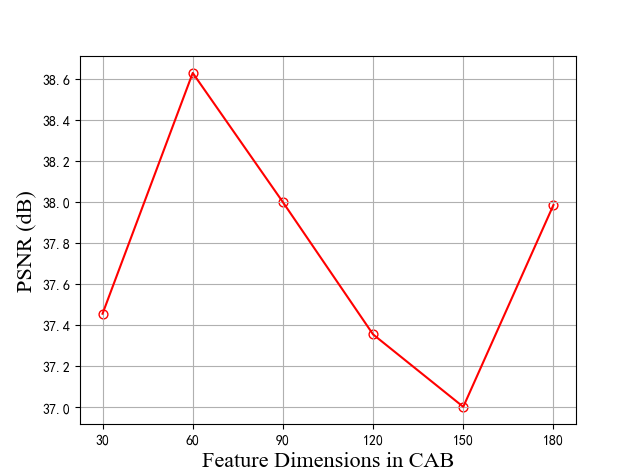}\label{fig:ablation_channelnumber}}\vspace{-0.1cm}
		\subfloat{\includegraphics[width=0.32\textwidth]{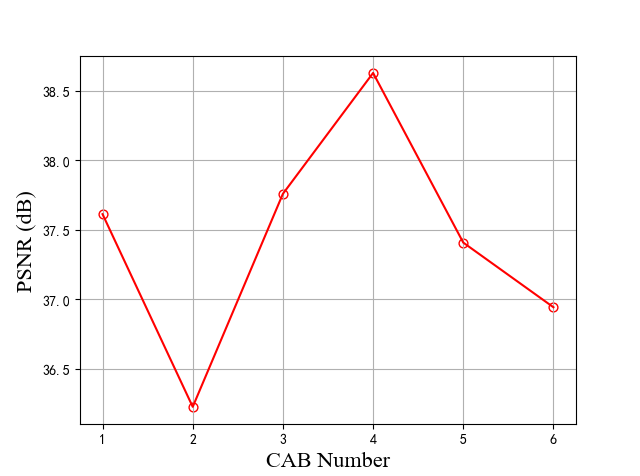}\label{fig:ablation_RDECABnumber}} \vspace{-0.1cm}
		\subfloat{\includegraphics[width=0.32\textwidth]{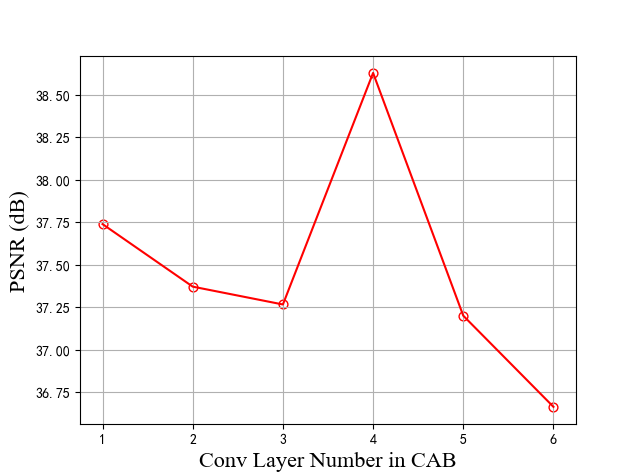}\label{fig:ablation_Convlayernumber}} \vspace{-0.1cm}
	\end{center}
	\caption{Ablation study on different settings of HSIE. Results are generated from our LHSI dataset. (a) Dimensions of feature maps in CAB. (b) CAB number of the enlightening module. (c) Convolution layer number in CAB. }\medskip
	\vspace{-0.3cm}
	\label{fig:ablation}
\end{figure*}

Fig.~\ref{fig:ablation} investigates the impact of feature map dimensions in CAB, the CAB number of the enlightening module, and the convolution layer number of CAB in Fig.~\ref{fig:ablation_channelnumber}, Fig.~\ref{fig:ablation_RDECABnumber}, and Fig.~\ref{fig:ablation_Convlayernumber}, respectively. It can be seen that the MPSNR is positively correlated with feature map dimensions in CAB. While the feature map dimensions are set as 60, 90, and 180, the MPSNR reaches a comparable good performance. 
For the sake of model complexity, the feature map dimensions are set as 60 in the final model. For the CAB number, the MPSNR becomes saturated when it equals 4. Finally, Fig.~\ref{fig:ablation_Convlayernumber} shows that the proposed HSIE gains the best performance when the convolution layer count equals 4 in CAB.

\begin{figure}[htb]
	\centering
	\includegraphics[width=0.98\columnwidth]{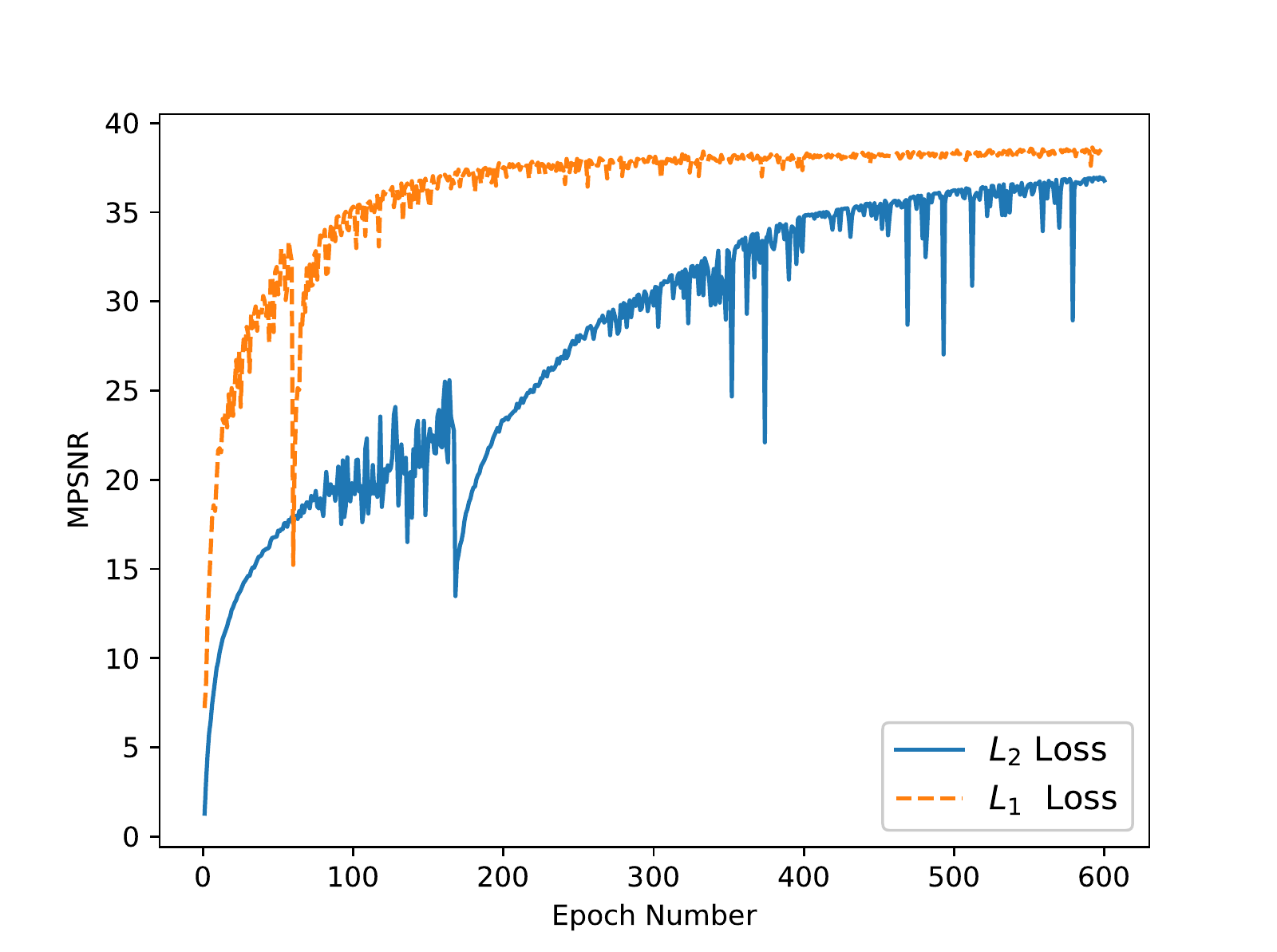}
	\caption{Investigation into different loss functions. The usage of the $L_1$ loss function results in a faster convergence speed and a more stable training process than the $L_2$ loss function.}
	\vspace{-0.3cm}
	\label{fig:ablation_loss_function}
\end{figure}

To evaluate the impact of different loss functions, the proposed HSIE are trained on indoor LHSI dataset using $L_1$ loss and $L_{2}$ loss respectively. Fig.~\ref{fig:ablation_loss_function} shows the MPSNR computed on the validation HSI at different epochs in training process. The $L 1$ loss function leads to a quicker convergence speed than the $L 2$ loss function, which is another reason we chose $L 1$ loss as the HSIE loss function. Furthermore, because the proposed HSIE is trained with the $L 1$ loss function rather than the $L 2$ loss function, the training process is more stable.

To ensure that the proposed HSIE is efficient, a computation complexity analysis experiment is conducted on the HSIE and corresponding deep-learning-based models. The model complexity is assessed by GFLOPs. As shown in Table~\ref{table:study_of_model_complexity}, four input spatial resolutions are exploited for the analysis, where the N.A. implies the  algorithm can not deal with the HSI of a certain spatial resolution on a single GPU with $24$G RAM. Results show that the proposed HSIE with a high-frequency refinement branch is more efficient than other methods on all four input spatial resolutions, and this superiority in computation efficiency becomes more and more obvious as the input spatial resolution increases.

\begin{table}[htb]
	\caption{A comparison of the model complexity of different deep-learning-based models. 
	}
	\vspace{-1.2em}
	\small
	\label{time_comparison}
	\begin{center}
		\begin{tabular}{p{2.3cm}|p{1cm}<{\centering}p{1cm}<{\centering}p{1cm}<{\centering}p{1cm}<{\centering}}
			\hline
			\multirow{2}{*}{Methods}  & \multicolumn{4}{c}{Complexity (GFLOPs)}  \\ \cline{2-5} 
			&$200\times200$&$384\times384$&$512\times512$&$1024\times1024$\\
			\hline
			ENCAM~\cite{encam} &105.8&N.A.&N.A.&N.A. \\
			3D-ADNet\cite{3d_danet}  &18.6&N.A.&N.A.&N.A. \\
			\hline
			\textbf{HSIE} (w/o high) &16.6&61.2&108.8&435.2\\
			\textbf{HSIE} (w/ high) &\textbf{13.1}&\textbf{48.4}&\textbf{85.9}&\textbf{343.9}\\
			\hline
		\end{tabular}
		\label{table:study_of_model_complexity}
	\end{center}
	\vspace{-0.6cm}
\end{table}

\begin{figure}[t]
	
	\begin{minipage}[b]{0.32\linewidth}
		\centering
		\centerline{\includegraphics[width=2.7cm]{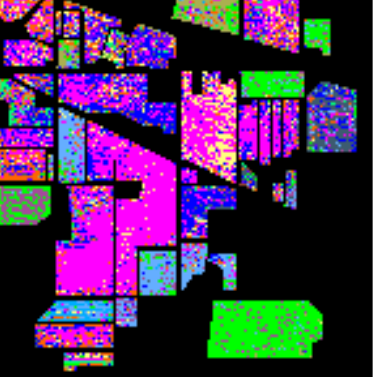}}
		\centerline{(a) Noisy}\medskip
	\end{minipage}
	\hfill
	\begin{minipage}[b]{0.32\linewidth}
		\centering
		\centerline{\includegraphics[width=2.7cm]{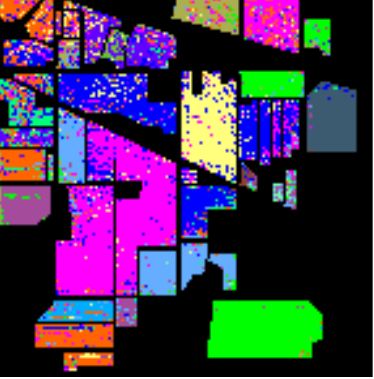}}
		\centerline{(b) LRMR}\medskip
	\end{minipage}
	\hfill
	\begin{minipage}[b]{0.32\linewidth}
		\centering
		\centerline{\includegraphics[width=2.7cm]{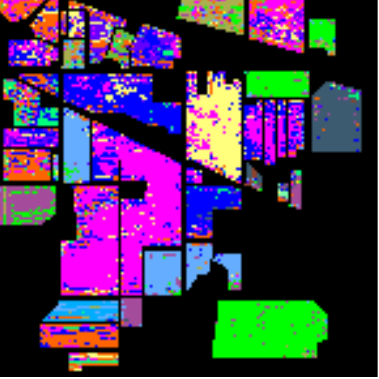}}
		\centerline{(c) BM4D}\medskip
	\end{minipage}
	\begin{minipage}[b]{0.32\linewidth}
		\centering
		\centerline{\includegraphics[width=2.7cm]{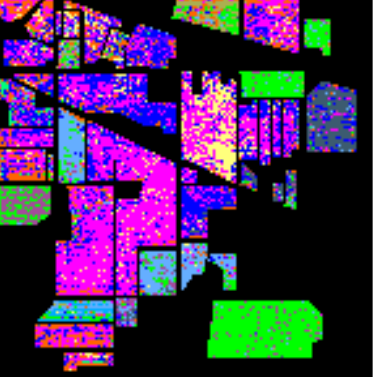}}
		\centerline{(d) HE}\medskip
	\end{minipage}
	\hfill
	\begin{minipage}[b]{0.32\linewidth}
		\centering
		\centerline{\includegraphics[width=2.7cm]{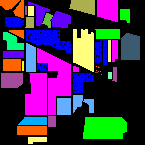}}
		\centerline{(e) Ground Truth}\medskip
	\end{minipage}
	\hfill
	\begin{minipage}[b]{0.32\linewidth}
		\centering
		\centerline{\includegraphics[width=2.7cm]{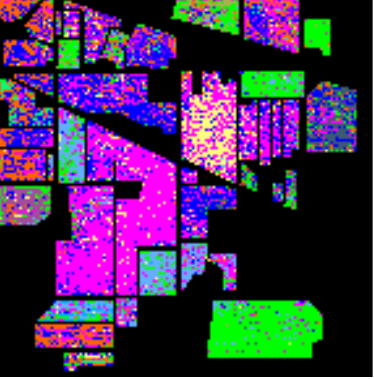}}
		\centerline{(f) CLAHE}\medskip
	\end{minipage}
	\begin{minipage}[b]{0.32\linewidth}
		\centering
		\centerline{\includegraphics[width=2.7cm]{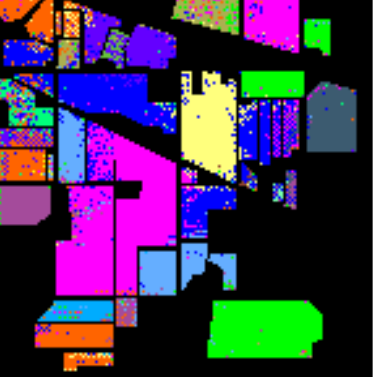}}
		\centerline{(g) MSR}\medskip
	\end{minipage}
	\hfill
	\begin{minipage}[b]{0.32\linewidth}
		\centering
		\centerline{\includegraphics[width=2.7cm]{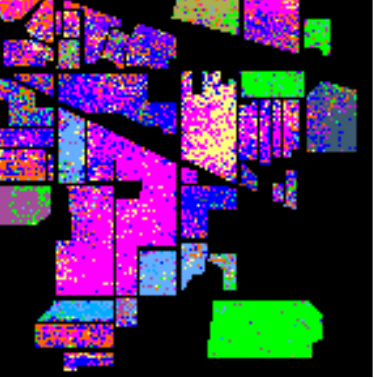}}
		\centerline{(h) 3D-ADNet}\medskip
	\end{minipage}
	\hfill
	\begin{minipage}[b]{0.32\linewidth}
		\centering
		\centerline{\includegraphics[width=2.7cm]{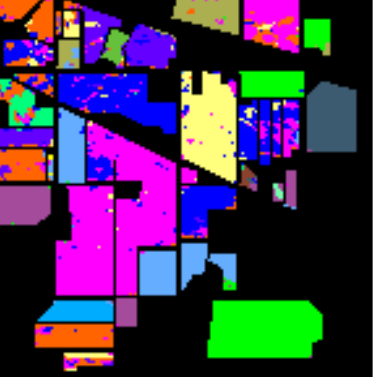}}
		\centerline{(i) HSIE (Ours)}\medskip
	\end{minipage}
	
	\vspace{-0.3cm}
	\caption
	{ Visualization of classification results using SVM on the well-known Indian Pine dataset. (a) Noisy. (b) LRMR. (c) BM4D. (d) HE. (e) Noisy-free. (f) CLAHE.
		(g) MSR. (h) 3D-ADNet. (i) HSIE (Ours).}\medskip
	\label{fig:indian_pine_svm_visual_result}  
	\vspace{-0.6cm}
\end{figure} 

\subsection{HSIs Classification Results}

\begin{table}[t]
	\centering
	\footnotesize
	\renewcommand{\arraystretch}{1.2}
	\caption{Classification performance using SVM on the enhanced low-light Indian Pines.}
	
	\begin{tabular}{l|c|c}
		\hline
		Models & OA$\uparrow$  & Kappa$\uparrow$ \\
		\hline
		Noisy                                & 0.4653                          & 0.3793                           \\
		LRMR\cite{LRMR}     & 0.7420                           & 0.7057                           \\
		BM4D\cite{bm4d}     & 0.6531                          & 0.6035                           \\
		HE\cite{gonzalez2009digital}                                   & 0.4914                          & 0.4072                           \\
		CLAHE\cite{park2008contrast}                                & 0.4767                          & 0.3885                           \\
		MSR\cite{jobson1997multiscale}                                  & 0.8212                          & 0.7957                           \\
		3D-ADNet\cite{3d_danet} & 0.5483                          & 0.4769                           \\ \hline
		\textbf{HSIE (Ours)} & \textbf{0.8770} & \textbf{0.8594} \\
		\hline
	\end{tabular}
	\smallskip
	
	\label{table:indian_pine_oa_kappa}
	\vspace{-0.6cm}
\end{table}	

As mentioned before, one purpose of enhancing low-light HSIs is to boost the performance of downstream tasks. We designed a classification task on the enhanced HSIs using the classic support vector machine (SVM) algorithm\cite{melgani2004classification} to certify the efficacy of the proposed HSIE. To carry out this experiment, we employ the classical remote sensing Indian Pine dataset, which is first darkened by a reversed HSIE network. This network is trained by exchanging the input and the label used for training the original HSIE. The enhanced results of varying methods on the darkened India Pine dataset are classified by SVM. We conduct this classification task without dimension reduction\cite{li2017locality}. The classification results on different enhanced HSIs and the original darkened India Pine data are presented in Fig.~\ref{fig:indian_pine_svm_visual_result}. The visualization of classification results on the original darkened HSI appears to have more corrupted and discontinuous areas, as illustrated in Fig.~\ref{fig:indian_pine_svm_visual_result} (a). After the original darkened HSI is enhanced by the proposed HSIE, the number of corrupted areas is reduced and more continuous areas are turned up (depicted in Fig.~\ref{fig:indian_pine_svm_visual_result} (i)). In this paper, we adopt the Kappa coefficient\cite{liu2019stacked} and overall accuracy (OA) to quantitatively evaluate the classification results, which are presented in Table~\ref{table:indian_pine_oa_kappa}. We can observe from Table~\ref{table:indian_pine_oa_kappa} that the Kappa and OA values of HSIE reach 0.8594 and 87.70\%,  respectively, which outperforms all the listed comparison methods.

\subsection{HSIs Denosing}

To reveal the applicability of our HSIE approach, we conduct an HSI denoising experiment on the remote sensing Washington DC Mall (WDC) dataset. We crop a cubic slice with shape $1000 \times 303 \times 191$ from the WDC dataset for training different methods and the rest for evaluating.

\begin{table}[htb]
	\small
	\setlength{\tabcolsep}{0.28mm}
	
	\caption{Denoising Performance of diverse algorithms on the Washington DC Mall dataset.}
	
	\begin{tabular}{crccccccc}
		
		\toprule[1pt]
		\begin{tabular}[c]{@{}c@{}}Noise\\ Level\end{tabular} & Metrics & BM4D & LRMR  & LRTA  & \begin{tabular}[c]{@{}c@{}}HSID\\ -CNN\end{tabular}  & \begin{tabular}[c]{@{}c@{}}3D-A\\ DNet\end{tabular} & ENCAM  & \begin{tabular}[c]{@{}c@{}}HSIE\\ (Ours)\end{tabular}    \\ \hline
		\multirow{3}{*}{5}                                  & MPSNR$\uparrow$ & 41.17 & 40.87 & 39.00 & 41.70 & \underline{42.08}                                               &      \textbf{42.72} & 34.15     \\ 
		& MSSIM$\uparrow$ & 0.996 & 0.995 & 0.993 & 0.996 & \underline{0.996}                                               &       \textbf{0.997} & 0.981    \\  
		& SAM$\downarrow$   & 1.933 & 2.276 & 2.701 & 1.832 & \underline{1.719}                                               &        \textbf{1.617} & 4.250   \\ \hline
		\multirow{3}{*}{25}                                  & MPSNR$\uparrow$ & 31.10 & 33.02 & 30.67 & 33.05 & 33.78                                               &     \textbf{33.90} & \underline{33.89}      \\  
		& MSSIM$\uparrow$ & 0.968 & 0.980 & 0.969 & 0.981 & 0.982                                              &        \textbf{0.983} & \underline{0.983}   \\ 
		& SAM$\downarrow$   & 5.051 & 4.609 & 5.796 & 4.264 & \underline{3.699}                                               &     \textbf{3.697} & 3.823      \\ \hline
		\multirow{3}{*}{50}                                  & MPSNR$\uparrow$ & 26.77 & 28.80 & 26.83 & 28.97 & 29.73                                               &    \textbf{30.04} & \underline{29.95}       \\ 
		& MSSIM$\uparrow$ & 0.918 & 0.953 & 0.925 & 0.954 & 0.959                                               & \textbf{0.964} & \underline{0.963}          \\ 
		& SAM$\downarrow$   & 7.141 & 6.800 & 7.500 & 6.220 & \textbf{5.015}                                               &\underline{5.049} & 5.441           \\ \hline
		\multirow{3}{*}{75}                                  & MPSNR$\uparrow$ & 24.29 & 26.30 & 24.68 & 26.75 & \underline{27.35}                                               & \textbf{27.75} & 27.33          \\  
		& MSSIM$\uparrow$ & 0.862 & 0.919 & 0.887 & 0.927 & 0.932                                               & \textbf{0.939} & \underline{0.936} \\  
		& SAM$\downarrow$   & 8.601 & 8.564 & 8.443 & 7.525 & \underline{6.138}                                               & \textbf{6.111} & 7.066 \\ \hline
		\multirow{3}{*}{100}                                  & MPSNR$\uparrow$ & 22.59 & 24.31 & 23.17 & 25.25 & \underline{25.74}                                               & \textbf{25.99} & 24.72 \\ 
		& MSSIM$\uparrow$ & 0.905 & 0.879 & 0.849 & 0.901 & \underline{0.906}                                               & \textbf{0.913} & 0.888 \\ 
		& SAM$\downarrow$   & 9.761 & 10.46 & 9.122 & 8.406 & \underline{7.322}                                               & \textbf{7.264}  & 9.577 \\ \bottomrule[1pt]
	\end{tabular}
	\vspace{-0.3cm}
	\label{table:comparision_on_wdcmall}
\end{table}

The results of the denoising performance on the WDC dataset are depicted in Table~\ref{table:comparision_on_wdcmall}. The top performance is bolded, while the second-best is underlined. Our HSIE delivers comparable results to the state-of-the-art method ENCAM, as shown by the results in Table~\ref{table:comparision_on_wdcmall}.

\section{CONCLUSION}  \label{sec:conclusion}
We offer a low-light hyperspectral image (LHSI) dataset in this work, which is created to support the evolution of low-light HSI enhancement approaches. We developed an end-to-end two-branch deep-learning-based network, dubbed HSIE, to raise the quality of low-light HSIs. According to the decomposition strategy of the Laplacian pyramid, we first decompose the input HSI into a low-resolution domain-specific component and a textural-related component. Then, the illumination enhancement branch is applied to boost the illumination with reduced resolution, while the lightweight high-frequency refinement branch aims to improve the textural details via a predicted mask. Substantial experiments on the LHSI dataset indicate that the proposed approach is capable of reaching promising performance in both evaluation metrics and visual effect, while at the same time suppressing noise and maintaining spectral fidelity. In addition, the accuracy of a downstream HSI classification task using SVM on enhanced low-light HSI demonstrated the benefits of HSIE as a low-light HSI preprocessing tool. This work offers many opportunities for future investigation into low-light HSI processing.

\bibliographystyle{IEEEtran}

\bibliography{IEEEabrv, ./reference.bib}

\vspace{0.5cm}

\begin{IEEEbiographynophoto}
	{Xuelong Li} is a Full Professor with School of Artificial Intelligence, OPtics and ElectroNics (iOPEN), Northwestern Polytechnical University, Xi’an 710072, P. R. China.
\end{IEEEbiographynophoto}
\vspace{-1cm}
\begin{IEEEbiography}
	[{\includegraphics[width=1in,height=1.25in,clip,keepaspectratio]{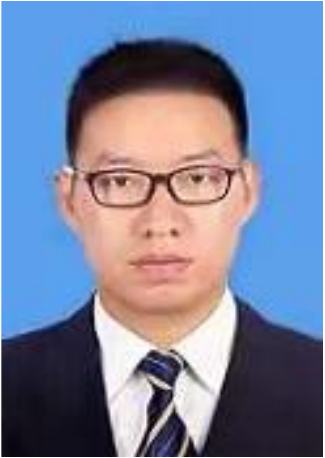}}]
	{Guanlin Li} is currently pursuing the Ph.D. degree with the School of Computer Science and School of Artificial Intelligence, OPtics and ElectroNics (iOPEN), Northwestern Polytechnical University, Xi’an, China. His research interests include generative image modeling, computer vision and machine learning.
\end{IEEEbiography}
\vspace{-1cm}
\begin{IEEEbiography}
	[{\includegraphics[width=1in,height=1.25in,clip,keepaspectratio]{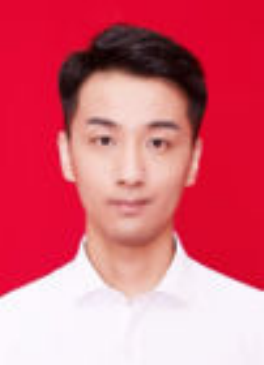}}]
	{Bin Zhao} is an Associate Professor with the School of Artificial Intelligence, OPtics and ElectroNics (iOPEN), Northwestern Polytechnical University, Xi’an 710072, P. R. China. His research interest is introducing physics models and cognitive science to artificial intelligence.
\end{IEEEbiography}

\end{document}